\documentclass[aps,pre,noshowpacs,onecolumn]{revtex4-1}

\usepackage{pstricks,pst-node}
\usepackage{amscd}
\usepackage{psfrag}
\usepackage{graphicx}
\usepackage{dcolumn}
\usepackage{bm}
\def\be{\begin{equation}}
\def\ee{\end{equation}}
\begin{document}

\title{Entropy production in systems with long range interactions}

\author{Renato Pakter}
\email{pakter@if.ufrgs.br}
\author{Yan Levin}
\email{levin@if.ufrgs.br}

\affiliation{Instituto de F\'{i}sica, UFRGS, Caixa Postal 15051, CEP 91501-970 Porto Alegre, Rio Grande do Sul, 
Brazil}

\begin{abstract}
On a fine grained scale the Gibbs entropy of an isolated  system remains constant throughout its dynamical evolution.  This is a consequence of Liouville's theorem for Hamiltonian
systems and {\it appears} to contradict the second law of thermodynamics. In reality, however, there is no problem since the thermodynamic entropy should be associated with the 
Boltzmann entropy, which for  non-equilibrium systems is different from Gibbs entropy.
The Boltzmann entropy accounts for the microstates which are not accessible from a given initial condition, but
are compatible with a given macrostate. In a sense the Boltzmann entropy is a coarse grained version of the 
Gibbs entropy and will not decrease during the dynamical evolution of a macroscopic system.  In this paper we will explore the entropy production for systems with long range interactions.  Unlike for short range systems, in the thermodynamic limit, the probability density function for these systems decouples into a product of one particle distribution functions and the coarse grained entropy can be calculated explicitly.  We find that the characteristic time for the entropy production scales with the number of particles as $N^\alpha$, with $\alpha > 0$, so that in the thermodynamic limit entropy production takes an infinite amount of time.

\end{abstract}

\maketitle

\section{Introduction}

Thermodynamics was a great triumph of the nineteenth century science. Clausius formulation of the second 
law of thermodynamics although deceptively simple: ``Entropy of the universe can not decrease", is profoundly complex. In apparent contradiction to the laws of Newtonian mechanics, which are invariant under the time reversal,  Clausius' second law introduces a preferred temporal direction -- the so called  ``arrow of time",
which points towards increasing entropy~\cite{Le93}.  
It took the genius of Boltzmann to reconcile the reversibility of microscopic dynamics with the irreversibility of thermodynamic systems.  Boltzmann associated the thermodynamic entropy of a $N$ particle system 
with the logarithm of the 
volume of a hypersurface in the 6N dimensional phase space compatible with the system's  macrostate. Note that 
there is a huge number of microstates -- parametrized by the positions and momentums of individual  particles -- compatible with a given macrostate, parametrized by the macroscopic observables. 
In this way Boltzmann cleverly escaped the constraints of the Liouville's theorem which 
establishes that a large portion of the phase space is actually inaccessible from a given initial 
condition, while being  {\it compatible} with the {\it evolving} macrostate. 
The increase of the Boltzmann entropy, therefore,
corresponds to the increase of the phase space {\it compatible} with the evolution of the macrostate. 
Clearly, if during the dynamical evolution all the velocities of the particles are reversed, the system will return to its initial state.   The probability of such event occurring spontaneously, however, is extremely small  -- it corresponds to one microstate 
among a huge number that are compatible with a given macrostate -- and, therefore, will not be observed in a {\it macroscopic} system for a reasonable experimental time scale.  For a system in thermodynamic equilibrium 
the Boltzmann thermodynamic entropy is $S_B=k_B \ln \Omega$, where $\Omega$ is the phase space volume of a constant energy hypersuface. 

While the calculation of the Boltzmann entropy is fairly straightforward for systems in 
thermodynamic equilibrium, this is not so 
for systems far from equilibrium for which it is very difficult to innumerate the microstates compatible
with a non-equilibrium macrostate.  In this case one might think that the Gibbs entropy is a better candidate 
for studying non-equilibrium systems~\cite{Gibbs}.  The Gibbs entropy is defined in terms of the probability density function $f({\bf q}^N, {\bf p}^N)$ which specifies the probability of finding $N$ particles at a given position 
in the phase space,  
\begin{equation}
S_G=-k_B \int d {\bf q}^N d {\bf p}^N f({\bf q}^N, {\bf p}^N, t) \ln f({\bf q}^N, {\bf p}^N, t) \,. \label{gib}
\end{equation} 
The dynamical evolution of the distribution function is governed by the Liouville equation.
Since all the microstates are assumed to be equally probable, the probability distribution density function for an equilibrium  system is $f({\bf q}^N, {\bf p}^N)=1/\Omega$ 
and the Gibbs entropy reduces to the Boltzmann entropy. 
Starting from a non-equilibrium state, however, the evolution of the Gibbs and Boltzmann entropies is very distinct.  The Liouville equation shows that the probability density evolves as an incompressible fluid, 
and is conserved along the flow.  This means that the phase space integral of any local function  of the distribution $h(f)$ will be preserved by the Hamiltonian dynamics.  This implies that the Gibbs entropy of the 
evolving system remains the same for all time!  
Gibbs was aware of this difficulty and suggested that to reconcile 
his entropy with the Clausius thermodynamic entropy one should use a coarse grained probability density
\begin{equation}\label{fbar}
\bar f(\mathbf{q}^N, \mathbf{p}^N,t)=\frac{1}{(\Delta p \Delta q)^{N d}} \int_{\Delta p^N, \Delta q^N} f(\mathbf{q'}^N, \mathbf{p'}^N,t)d\mathbf{q'}^N d\mathbf{p'}^N \,,
\end{equation}
where $d$ is the dimensionality of the configuration space, 
in Eq. (\ref{gib}), resulting in a coarse grained Gibbs entropy $\bar S_G$.  
The coarse graining procedure is similar to the Boltzmann counting of the microstates
which are not accessible from a given initial condition, but are compatible with a given macrostate. It accounts for the gradual loss of information as the probability density function filaments over the available phase space. In any simulation or experiment with
a macroscopic system one rapidly runs into a resolution limit beyond which one has no longer any precision to
specify the fine structure of the distribution function.  This is similar to a tightly wound spool of wire, which
from a distance appears to be a ball of uniform density.  

\section{Systems with long range interactions}

The difficulty of calculating $\bar S_G$ 
is that for usual thermodynamic systems of particles interacting by short range potentials
the coarse graining has to be performed in the full 2dN dimensional $\Gamma$ phase space and is very difficult to implement in practice. 
We note, however, that for systems with long range interactions for which the interaction potential decays with
distance slower than the dimensionality of the embedding configuration space, {\it in the thermodynamic limit}, the probability density function decouples into a product of one particle distribution functions~\cite{Br77},
\begin{equation}\label{dis}
f({\bf q}^N, {\bf p}^N, t)=\prod_{i=1}^N f_1({\bf q}_i, {\bf p}_i, t) \,,
\end{equation}
such that, 
\begin{equation}\label{norm}
\int f_1({\bf q}, {\bf p}, t)  d {\bf q} d {\bf p} =1 \,.
\end{equation}
This means that the full $\Gamma$ phase space 
effectively collapses to a 2d dimensional reduced $\mu$ phase space.  
The Liouville equation which governs the evolution of the probability density function 
reduces to a much simpler Vlasov equation
for one particle distribution function
\begin{equation}\label{eq:vlasov}
\left(\frac{\partial}{\partial t}+\mathbf{p}\cdot\frac{\partial}{\partial \mathbf{q}}-\frac{\partial \psi (\mathbf{q},t)}{\partial \mathbf{q}}\cdot
\frac{\partial}{\partial \mathbf{p}}\right)f_1(\mathbf{q},\mathbf{p},t)=0.
\end{equation}
where $\psi(\mathbf{q},t)$ is the mean field potential felt by a particle located at position $\mathbf{q}$.
Furthermore, for long range interacting systems the Gibbs entropy simplifies to
\begin{equation}
S_G=-k_B N\int d {\bf q} d {\bf p} f_1({\bf q}, {\bf p}, t) \ln f_1({\bf q}, {\bf p}, t)\,, \label{gib1}
\end{equation} 
which is also conserved by the Vlasov flow.  The advantage of working with the long range systems
is that the Gibbs coarse graining procedure is much easier to implement in the reduced $\mu$ phase spaces than
in the full $\Gamma$ phase space.  The simplest way to calculate $\bar S_G$ is using a 
non-parametric entropy estimator~\cite{Do58,KoLe87,SiMi03}.
The idea is to coarse grain the distribution function on the scale of the nearest neighbor distance between the particles.  Such coarse graining is scale-free and is easy to implement in numerical molecular dynamics simulations. The coarse
grained entropy can then be written as
\begin{equation}
\bar S_G=k_B \sum_i^N \ln \left(N r_i^{2d} V_{2d} \right) + N \gamma\,,  \label{gib2}
\end{equation}
where $r_i$ is the Euclidean distance in the $\mu$ phase space between the particle $i$ and its nearest neighbor, $V_{2d}=\pi^d/\Gamma(d+1)$ is the volume of a $2d$ dimensional unit sphere, and $\gamma= 0.57721...$ is the Euler-Mascheroni constant.

\section{Non-interacting particles}

\begin{figure}[!htb]
\begin{center}
\includegraphics[scale=.3]{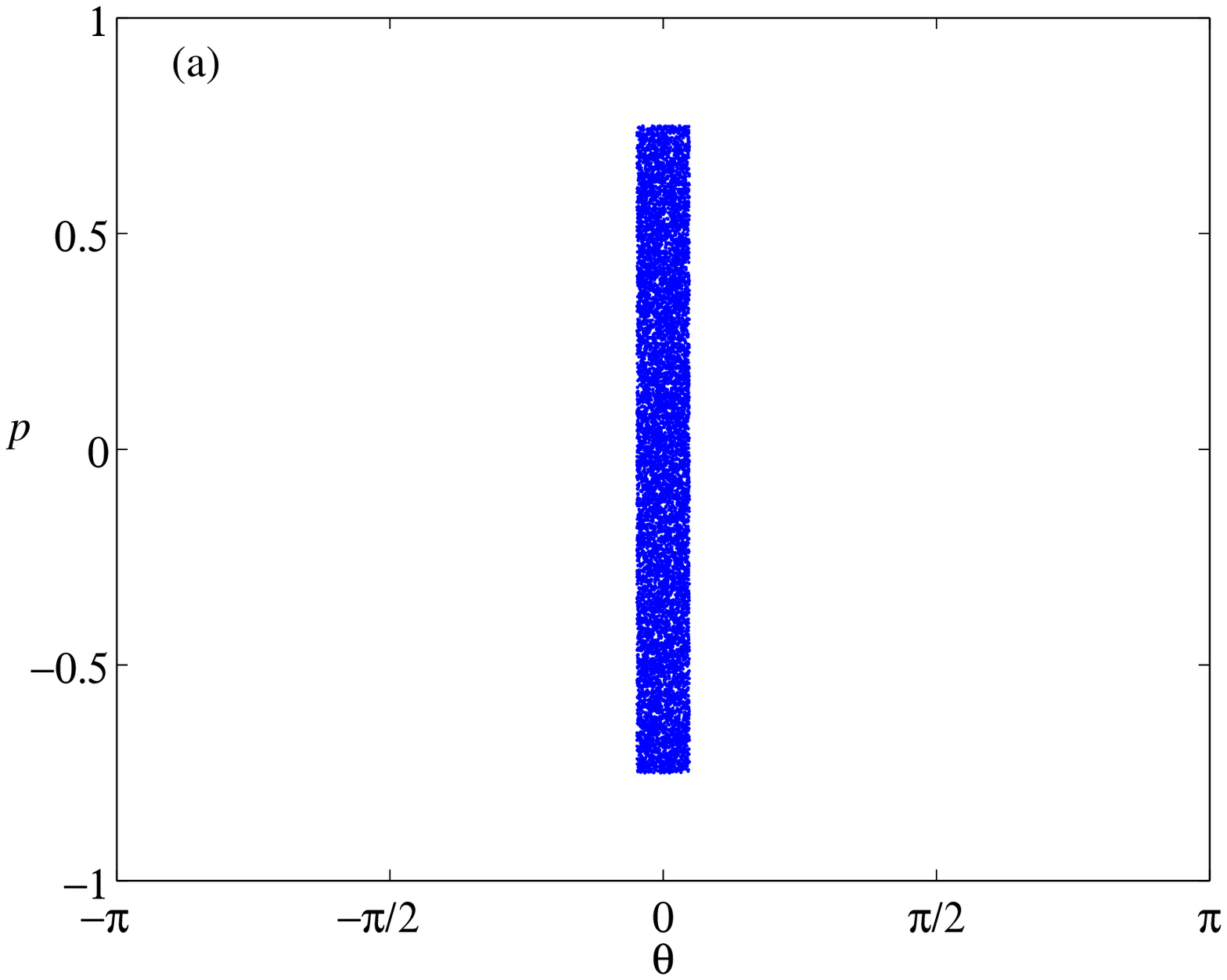}\hspace{1cm}
\includegraphics[scale=.3]{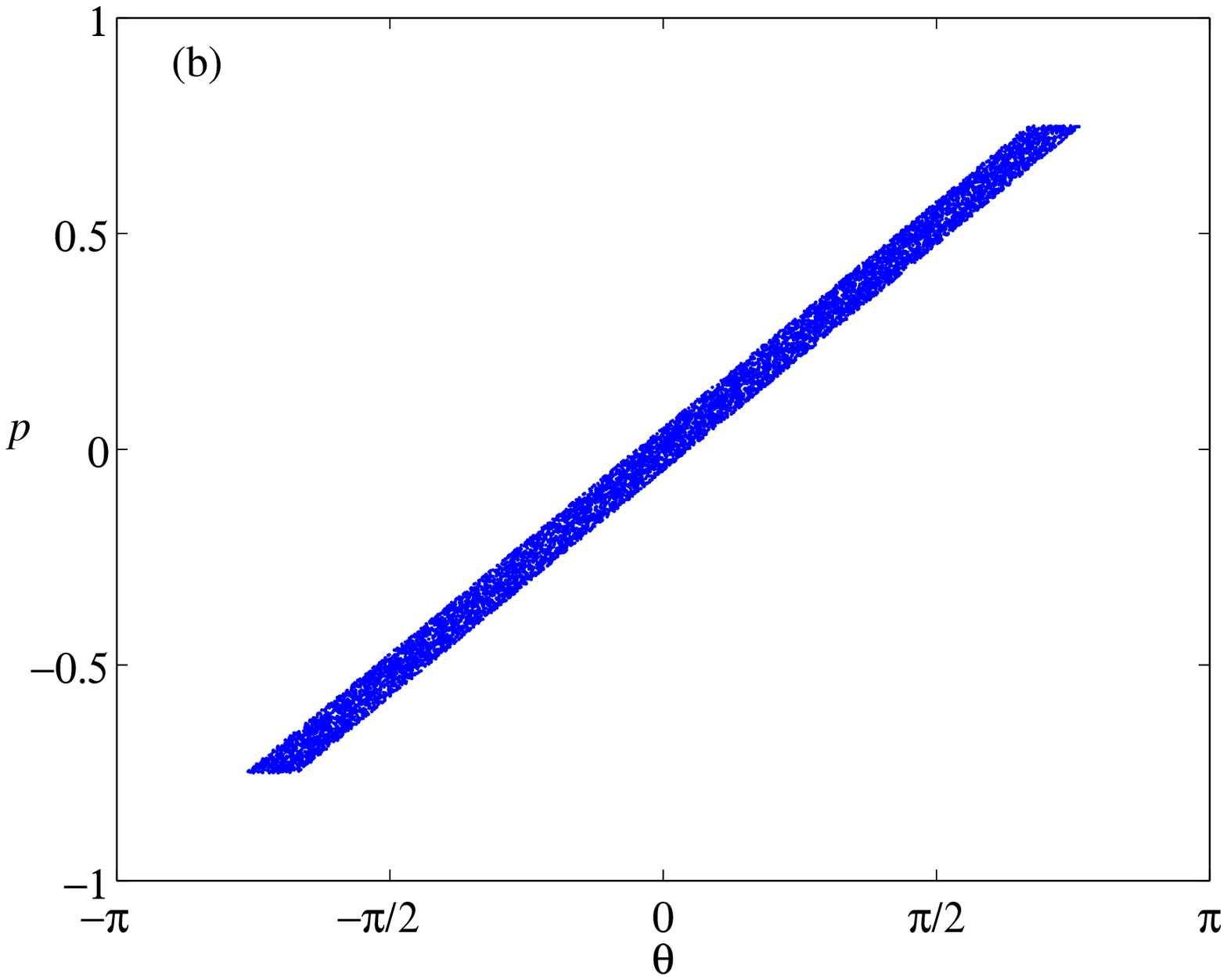}

\includegraphics[scale=.3]{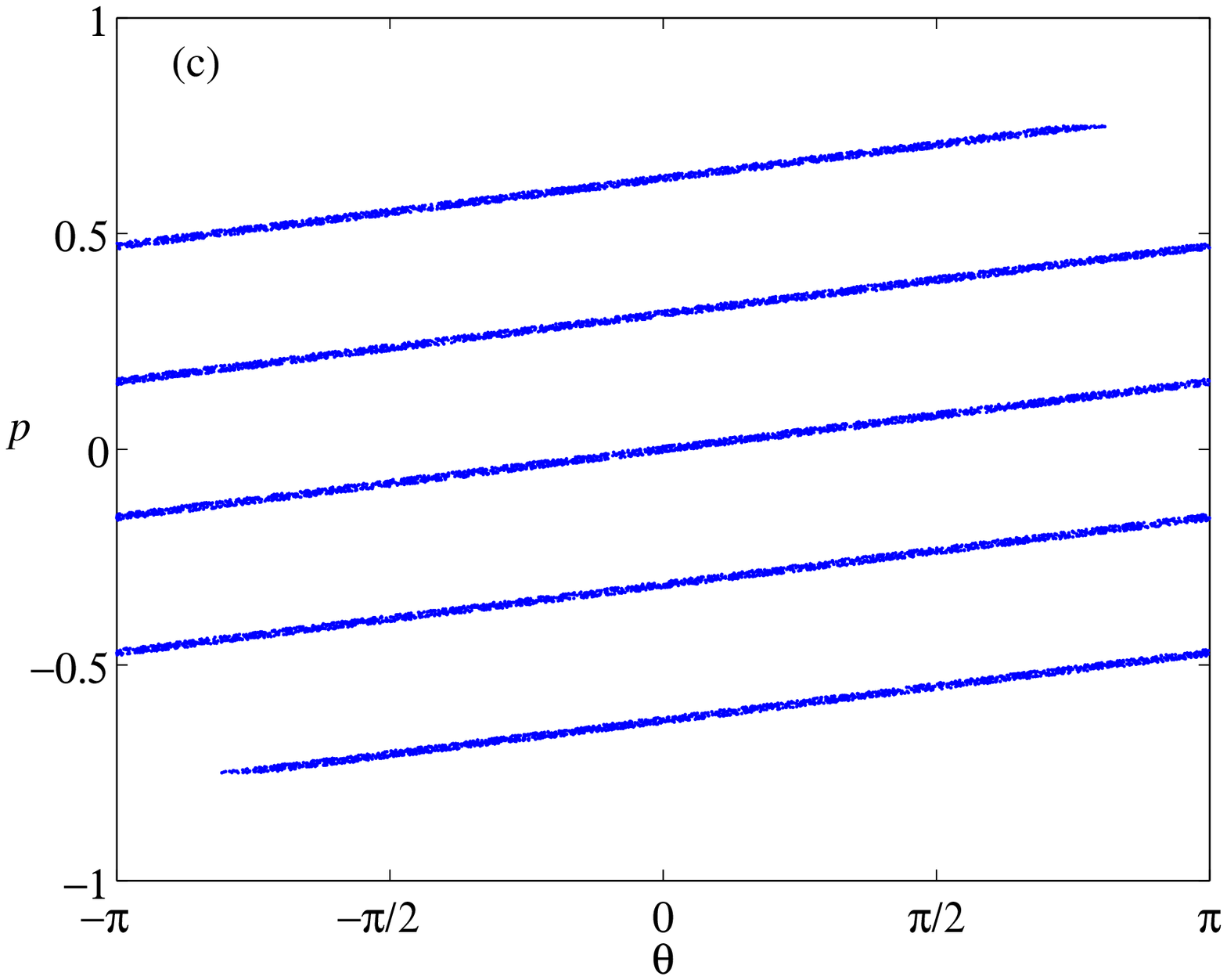}\hspace{1cm}
\includegraphics[scale=.3]{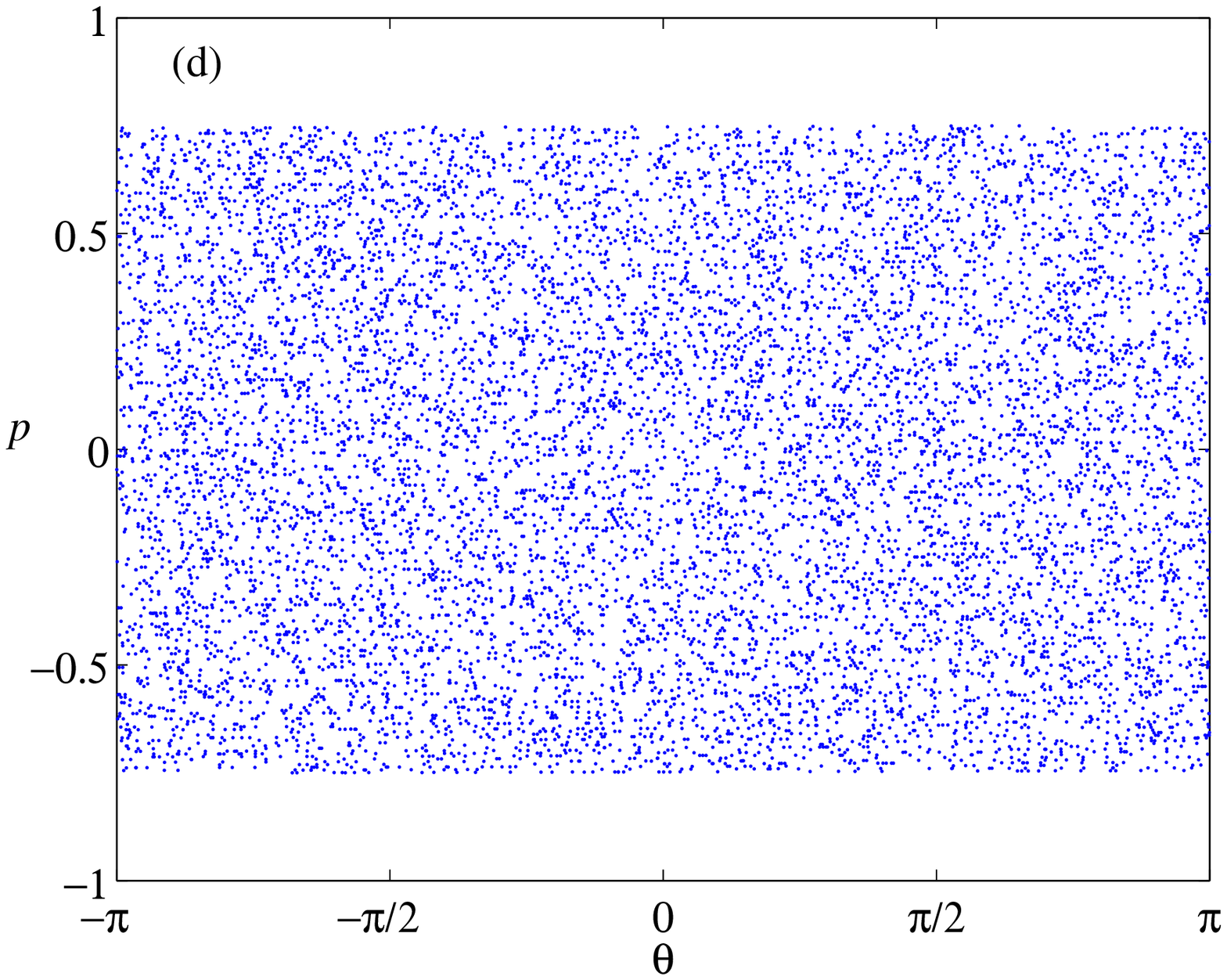}
\end{center}
\caption{Snapshots of $\mu$ phase space of a
noninteracting system: (a) $t=0.0$, (b) $t=3.0$, (c) $t=20.0$, and (d) $t=10^3$.}
\label{fig5}
\end{figure}

\begin{figure}[!htb]
\begin{center}
\includegraphics[scale=.3]{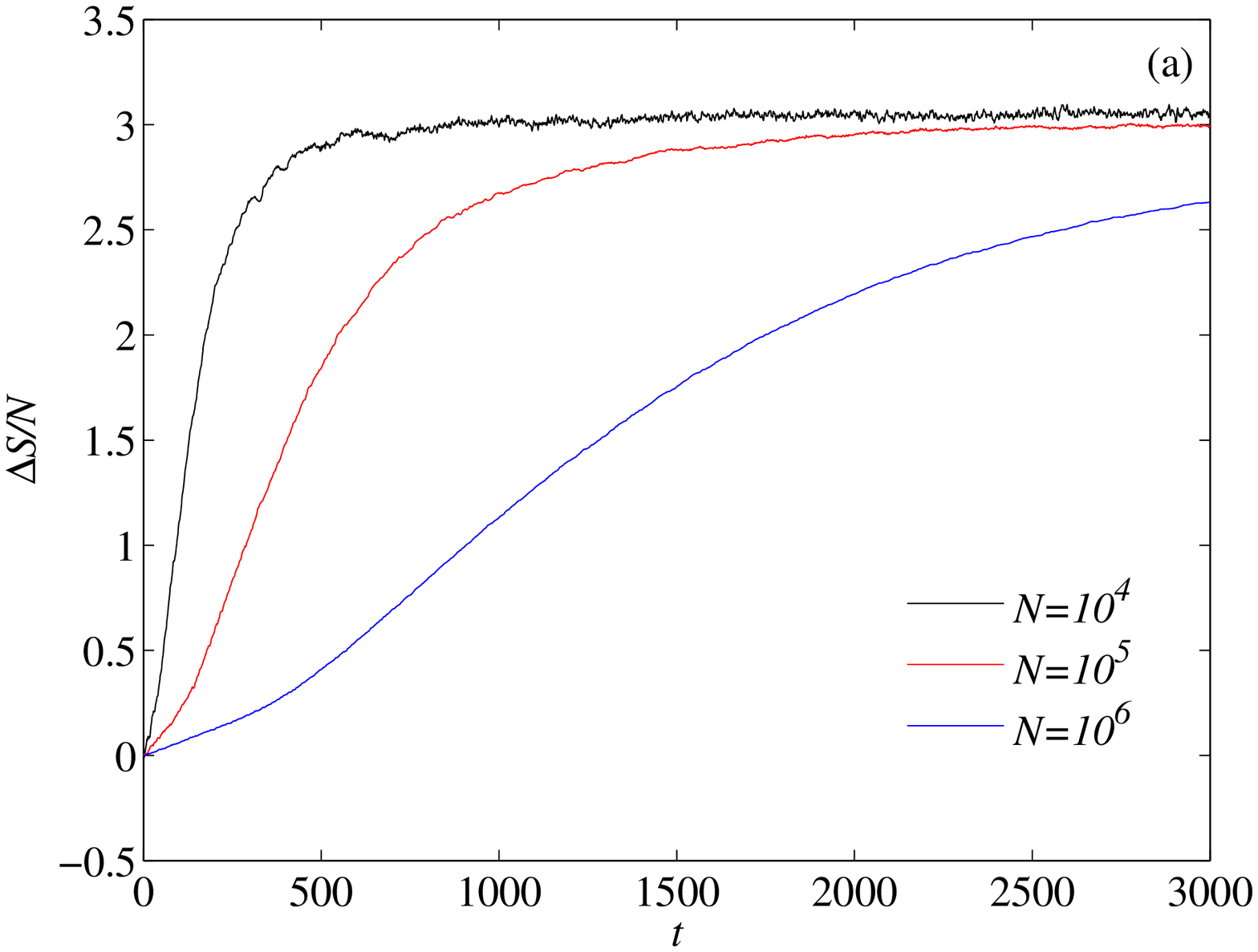}\hspace{1cm}
\includegraphics[scale=.3]{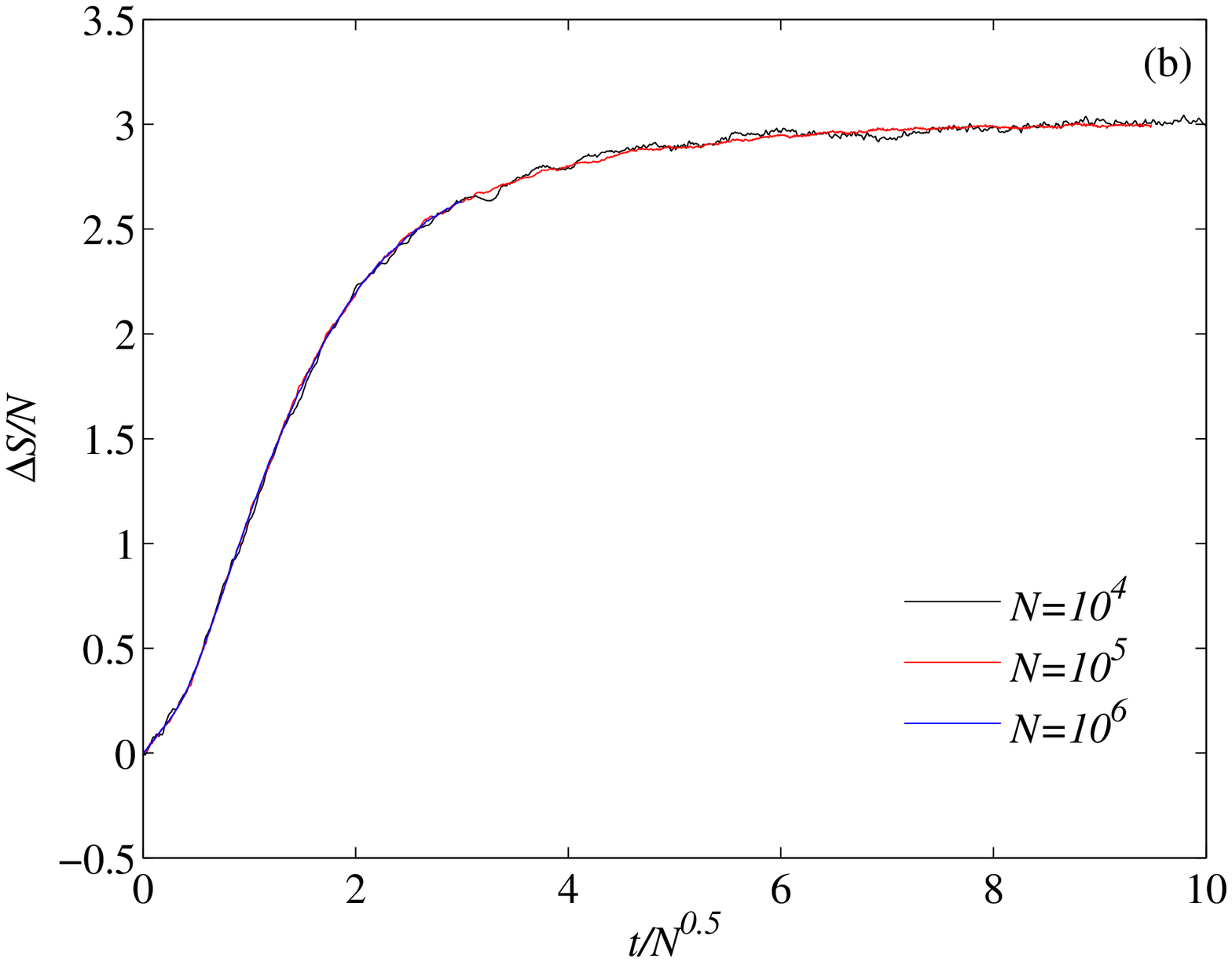}
\end{center}
\caption{Entropy production in a system of noninteracting particles: (a) dynamical time, (b) scaled time, showing a perfect data collapse.}
\label{fig1}
\end{figure}

We begin our study of the coarse grained entropy production with a system of non-interacting particles
confined to move on a unit ring,
\begin{equation}
H=\sum_{i=1}^N {p_i^2\over 2}\,.  \label{nonint}
\end{equation}
The Hamilton's equations of motion for this system take particularly simple form: $p_i(t)=p_i(0)$ and  $\theta_i(t)=\theta_i(0)+p_i(0) t$ where $\theta$ is defined between $-\pi$ and $\pi$. We start with a waterbag
initial distribution
\be
f_1(p,\theta, t=0)=\frac{1}{4 p_m \theta_m} \Theta(p_m-|p|)\Theta(\theta_m-|\theta|) \,,\label{wb}
\ee 
where $\Theta(x)$ is the Heaviside theta function.
The various snapshots of the $\mu$ phase space of $N$ particles undergoing the dynamics governed by the Hamiltonian (\ref{nonint}) are shown in Fig. \ref{fig5}.
We see the process of phase space filamentation and eventual loss of the resolution characteristic of
coarse grained entropy production. 
In Fig. \ref{fig1}a we plot the coarse grained 
entropy production per particle as
a function of time.  
The figure shows that the entropy production is strongly dependent on the number of particles
in the system.  A loss of the resolution is much faster with a fewer particles than when the system
is very large.  Rescaling the time with $N^{0.5}$ all the curves collapse onto one universal curve. 
Therefore, for this non-interacting particle model,  in the thermodynamic limit, entropy production will 
require infinite amount of time, and the fine structure resolution of the phase space will persist up to 
microscopic length scales.  

We stress that in order to explore the finite size scaling of the entropy production time it is necessary 
to have a sufficiently accurate estimate of the coarse grained entropy {\it i.e} sufficiently good resolution of the phase space.  Expression, Eq. (\ref{gib2}), meets this
requirement.  To check its accuracy we can compare the calculated coarse grained entropy obtained using Eq. (\ref{gib2}) for the final stationary state to which a non-interacting system relaxes, with the exact
value.  Since the filamentation process leads to a uniform occupation of the phase space, see Fig. \ref{fig5}d, in the $t \rightarrow \infty$ limit, the coarse grained
distribution function  evolves to $f_1(p,\theta, t=\infty)=1/4 \pi p_m$ and  the coarse grained entropy of the stationary state {\it in the thermodynamic limit} is exactly $\bar S_G=N \ln (4 \pi p_m)$. 
Even with a relatively small number of particles $N=10^4$, the non-parametric estimator
leads to an error in $\bar S_G$ of less than half of a percent.  
To obtain such a small error using a brute force calculation
of the distribution function requires  $10^6$ phase space cells and $10^8$ particles, which makes such methods impractical for interacting systems.  The difficulty is that to get a sufficiently accurate representation of the filamentation process requires a large number of phase space cells.  On the other hand, to have 
a reasonable statistics, demands a sufficiently large particle occupation of each of these cells, forcing us to work with a very large number of particles.  To avoid these problems in the present paper we use the non-parametric estimator Eq. (\ref{gib2}) to explore the finite size scaling of the entropy production time.

\section{Entropy production in Hamiltonian Mean Field (HMF) Model }

Unlike systems of particles interacting by short range forces,  systems with  long range (LR) interactions, such as magnetically confined one component plasmas or self gravitating systems, in the infinite $N$ limit do not relax to thermodynamic equilibrium but become trapped in non-equilibrium quasi stationary
states (QSS)~\cite{LePa14,CaDa09}.  
This means that the coarse grained entropy of these systems does not reach the maximum entropy state
of thermodynamic equilibrium.  Nevertheless, the second law of 
thermodynamics requires that starting from an initial configuration far from QSS, 
the coarse grained entropy 
must increase. To quantitatively calculate this entropy production 
we study a paradigmatic Hamiltonian Mean Field (HMF) model of a system 
with LR interactions.
The HMF model consists of $N$, $XY$ interacting  spins,  whose dynamics
is governed by the Hamiltonian
\be
H=\sum_{i=1}^N {p_i^2\over 2}+{1\over 2 N}\sum _{i,j=1}^N [1-\cos(\theta_i-\theta_j)],
\ee
where angle $\theta_i$ is the orientation of the $i$th spin 
and $p_i$ is its conjugate momentum~\cite{AnFa07,AnFa07b,AnCa07,PaLe11}. 
Alternatively the model can be thought of as particles confined to move on a unit ring with
all the particles interacting by a cosine potential. Note that since each spin interacts with
every other spin by the potential which only depends on the relative orientation and not on 
spin separation, the interaction is infinitely long range.

\begin{figure}[!htb]
\begin{center}
\includegraphics[scale=.3]{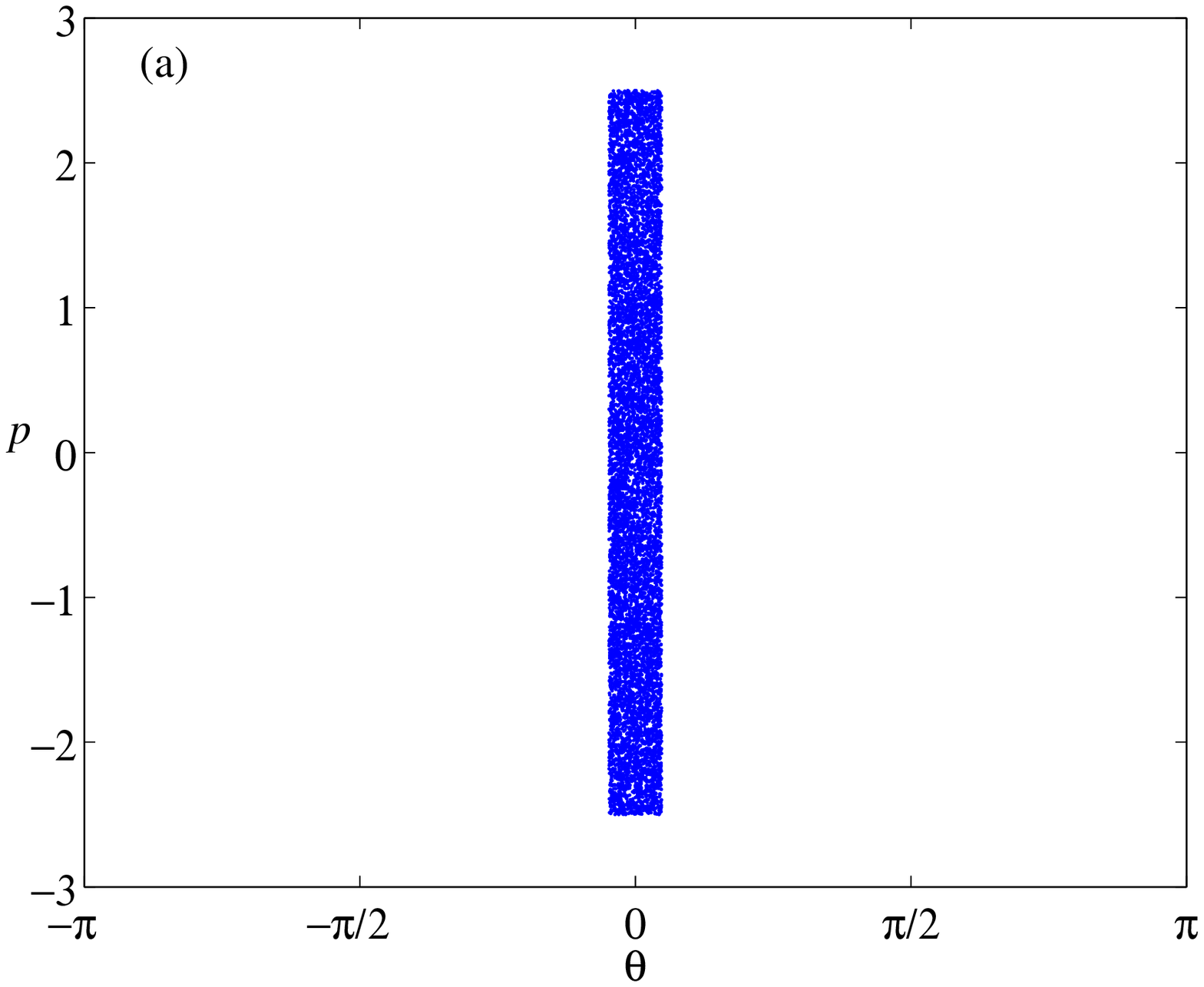}\hspace{1cm}
\includegraphics[scale=.3]{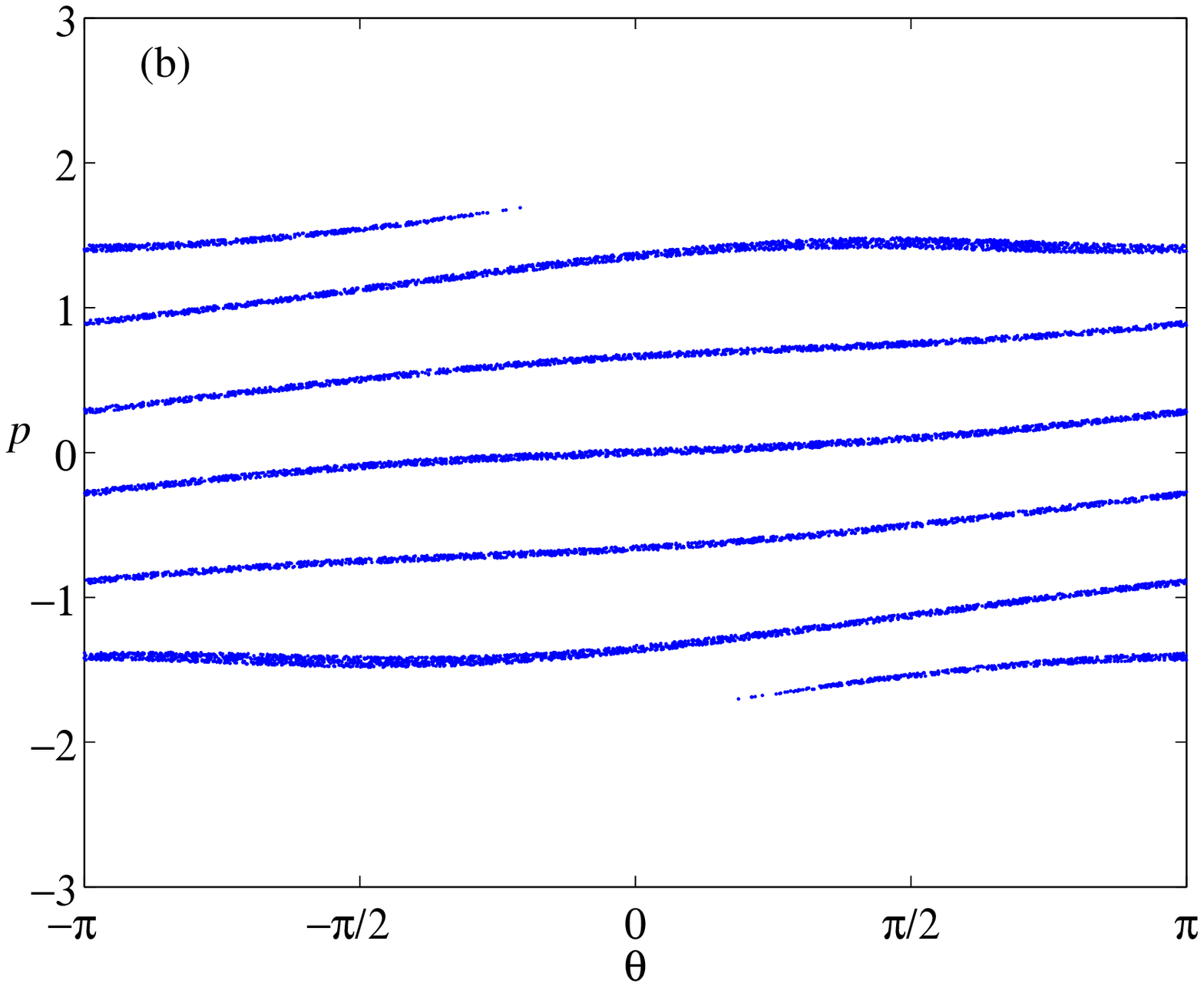}

\includegraphics[scale=.3]{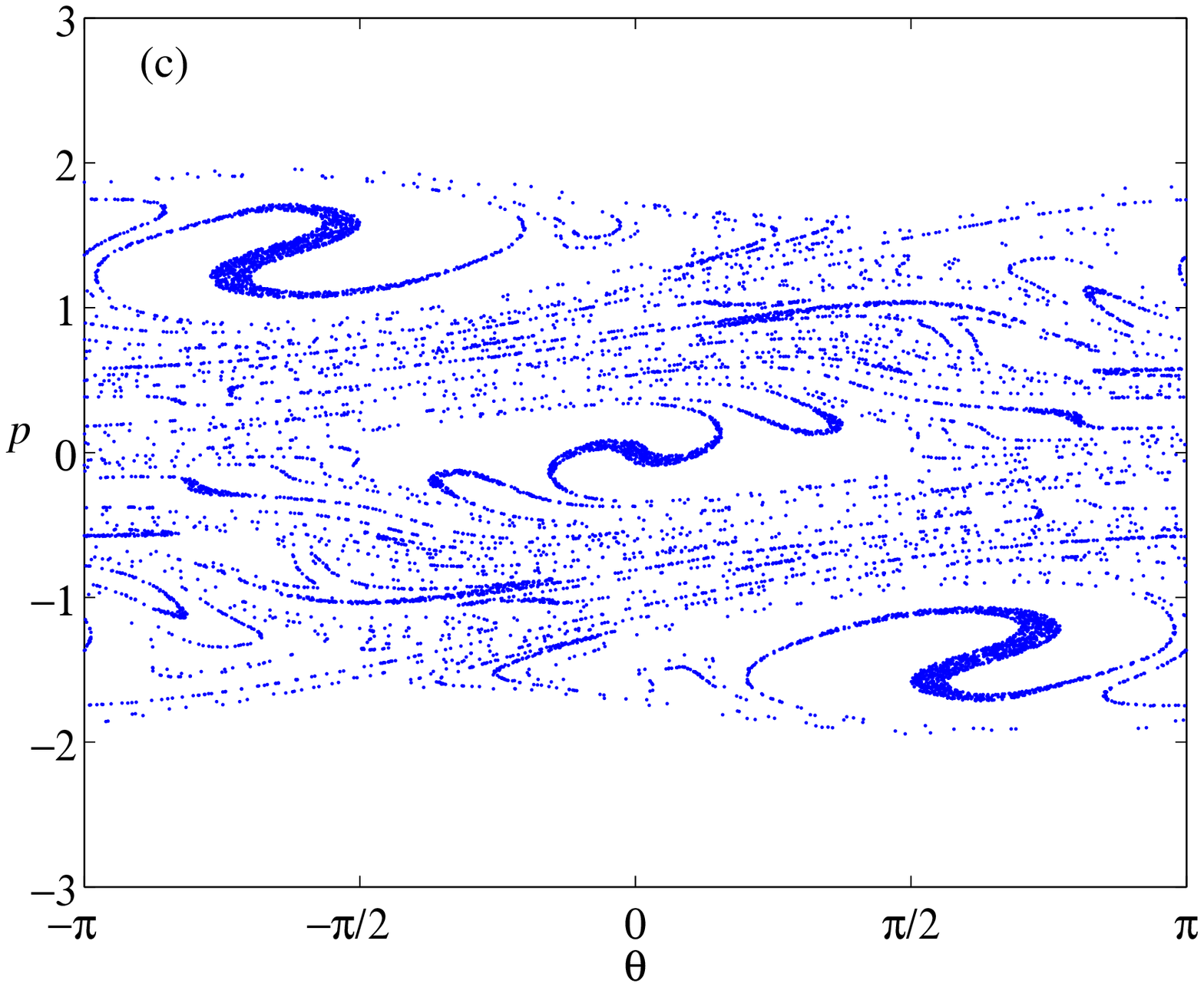}\hspace{1cm}
\includegraphics[scale=.3]{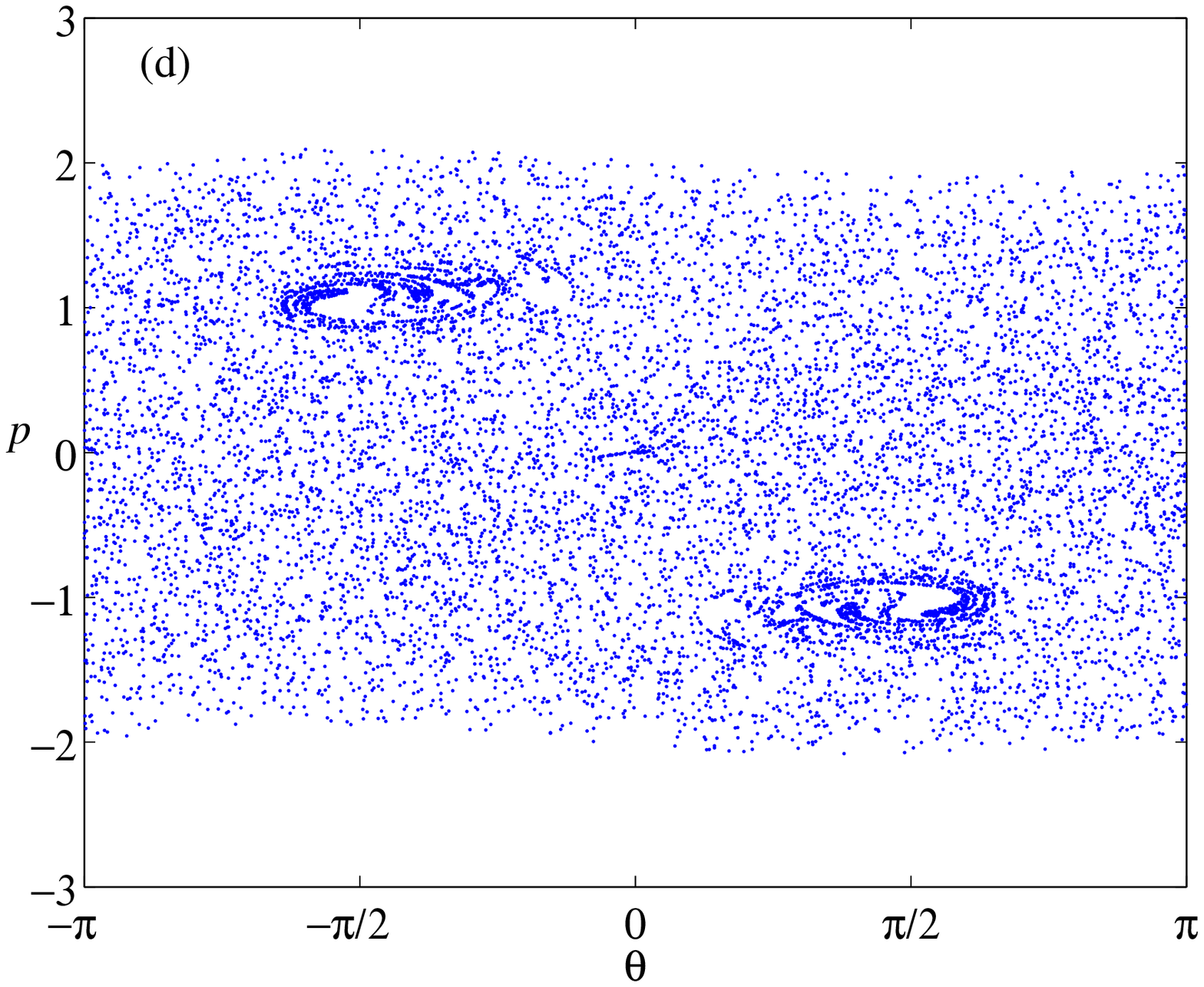}
\end{center}
\caption{Snapshots of the evolution of the $\mu$ phase space of the HMF model starting from a non-equilibrium magnetized initial condition of a waterbag form ($\theta_m=0.3$, $p_m=2.50$).  The final QSS is unmagnetized: (a) $t=0.0$, (b) $t=100.0$, (c) $t=500.0$, and (d) $t=10^4$.}
\label{fig6}
\end{figure}

\begin{figure}[!htb]
\begin{center}
\includegraphics[scale=.3]{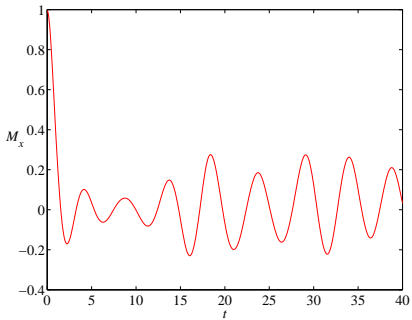}\hspace{1cm}
\end{center}
\caption{$M_x$ as a function of time for an initially magnetized state.  
The final QSS to which the system evolves is unmagnetized -- temporal average of $M_x(t)$ is zero.}
\label{fig9}
\end{figure}

The {\it macroscopic} behavior of the system is characterized by the magnetization vector
${\bf M}=(M_x,M_y)$, where $M_x\equiv \langle \cos\theta \rangle$, 
$M_y\equiv \langle \sin\theta \rangle$, and $\langle\cdots\rangle$ stands for the average
over all particles. The modulus, $M=|{\bf M}|$ serves as the order parameter which measures the
coherence of the spin angular distribution: for $M=0$ we have a completely
disordered state, whereas for finite $M$ there is some degree of order (coherence). 
Hamilton's equations of motion for each spin can be expressed in terms of 
the total magnetization and reduce to a second order differential equation, 
\be
\ddot\theta_i=-M_x\sin\theta_i+M_y\cos \theta_i \,.
\label{evol}
\ee
Because of the symmetry of $f_1(p,\theta, t=0)$ with
respect to $\theta=0$, in
the thermodynamic limit $M_y(t)=0$ throughout the evolution, so that the macroscopic dynamics is
completely determined by $M_x(t)$, which becomes the order parameter~\cite{PaLe11}.
In Fig. (\ref{fig6}) we show the evolution of the $\mu$ phase space of the HMF model starting from an initial
distribution of a waterbag form, Eq. (\ref{wb}), with finite initial magnetization -- all spins are closely aligned but have a lot of kinetic energy. The final QSS to 
which the system relaxes is paramagnetic with the magnetization oscillating around zero, see Fig. \ref{fig9}.

\begin{figure}[!htb]
\begin{center}
\includegraphics[scale=.3]{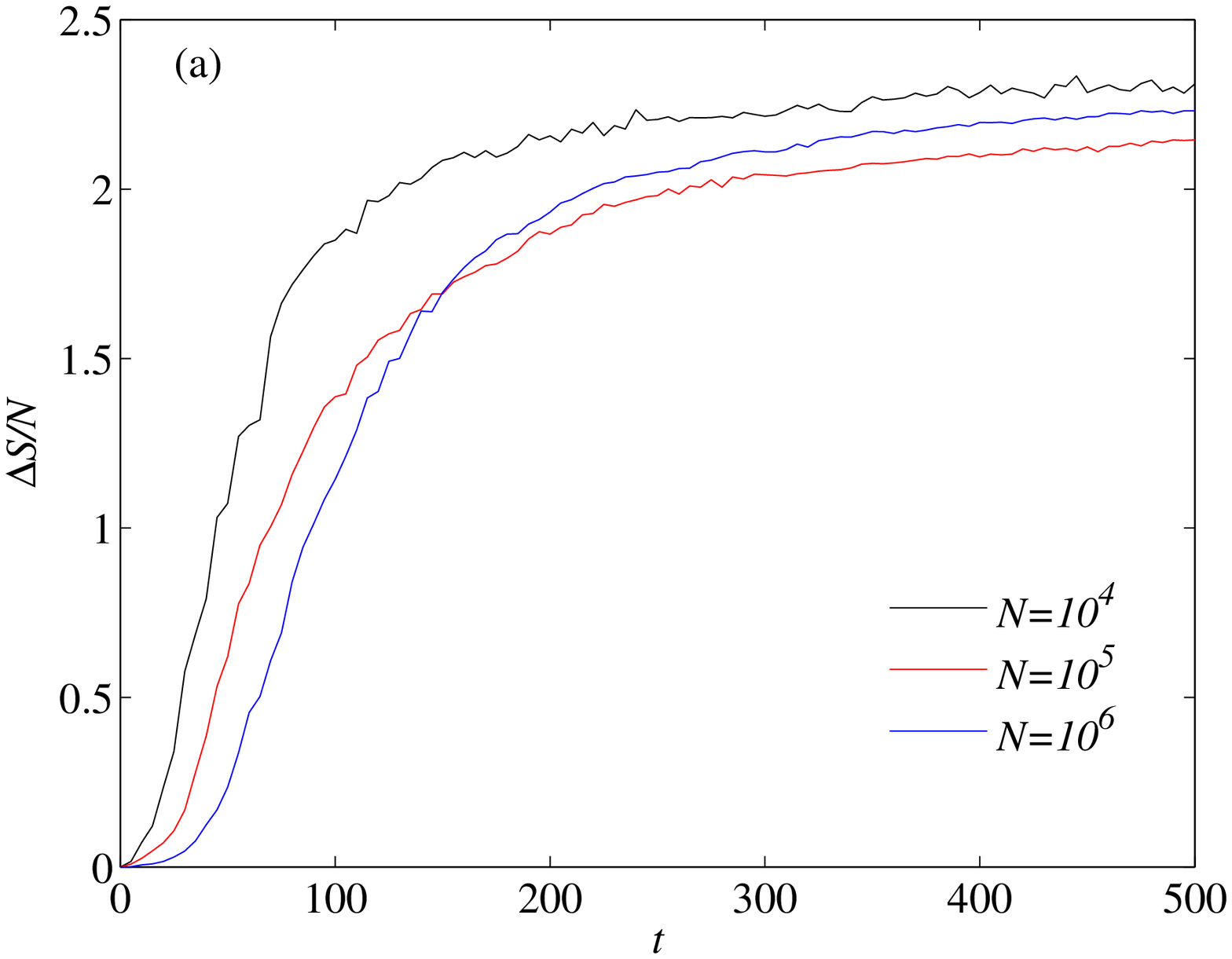}\hspace{1cm}
\includegraphics[scale=.3]{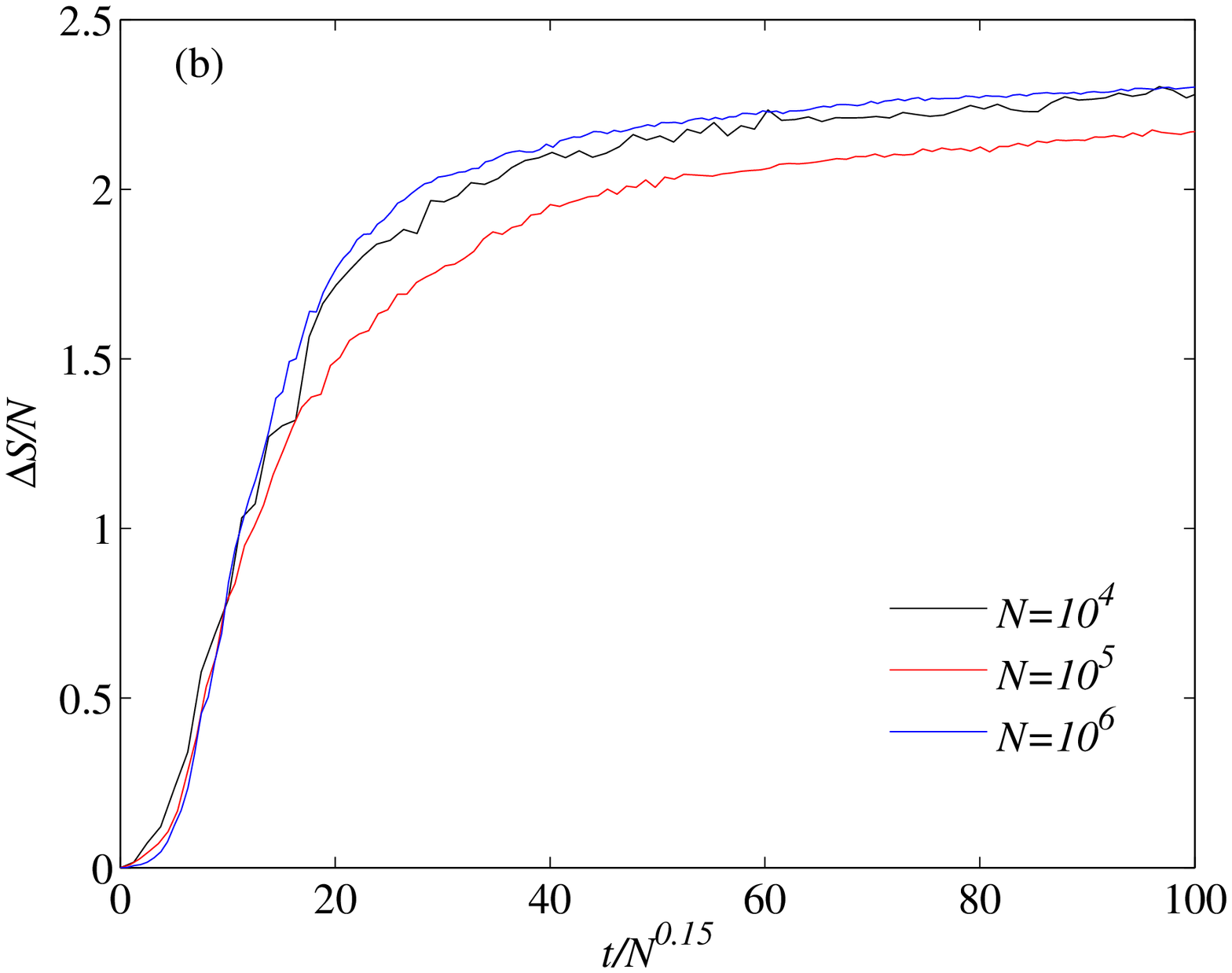}
\end{center}
\caption{ Entropy production in the HMF model starting from a non-equilibrium magnetized initial condition of a waterbag form ($\theta_m=0.3$, $p_m=2.50$) which then relaxes to an unmagnetized QSS: 
(a) dynamical time, (b) scaled time. }
\label{fig2}
\end{figure}

The evolution of the coarse grained entropy is shown in Fig. \ref{fig2}.
We see that the entropy production depends on the number of particles.  Scaling
time with $N^\alpha$, where $\alpha=0.15$, we obtain a good data collapse for the early times. 
The non-linear coupling between
the spins produced by the mean-field potential leads to a much faster entropy production than
for a system of non-interacting particles for which exponent was found to be $\alpha=0.5$.  The parametric  
resonances present in the dynamics of the HMF model lead to a more efficient phase space mixing and a faster
loss of fine-grain resolution than the one observed for non-interacting particles.

\begin{figure}[!htb]
\begin{center}
\includegraphics[scale=.3]{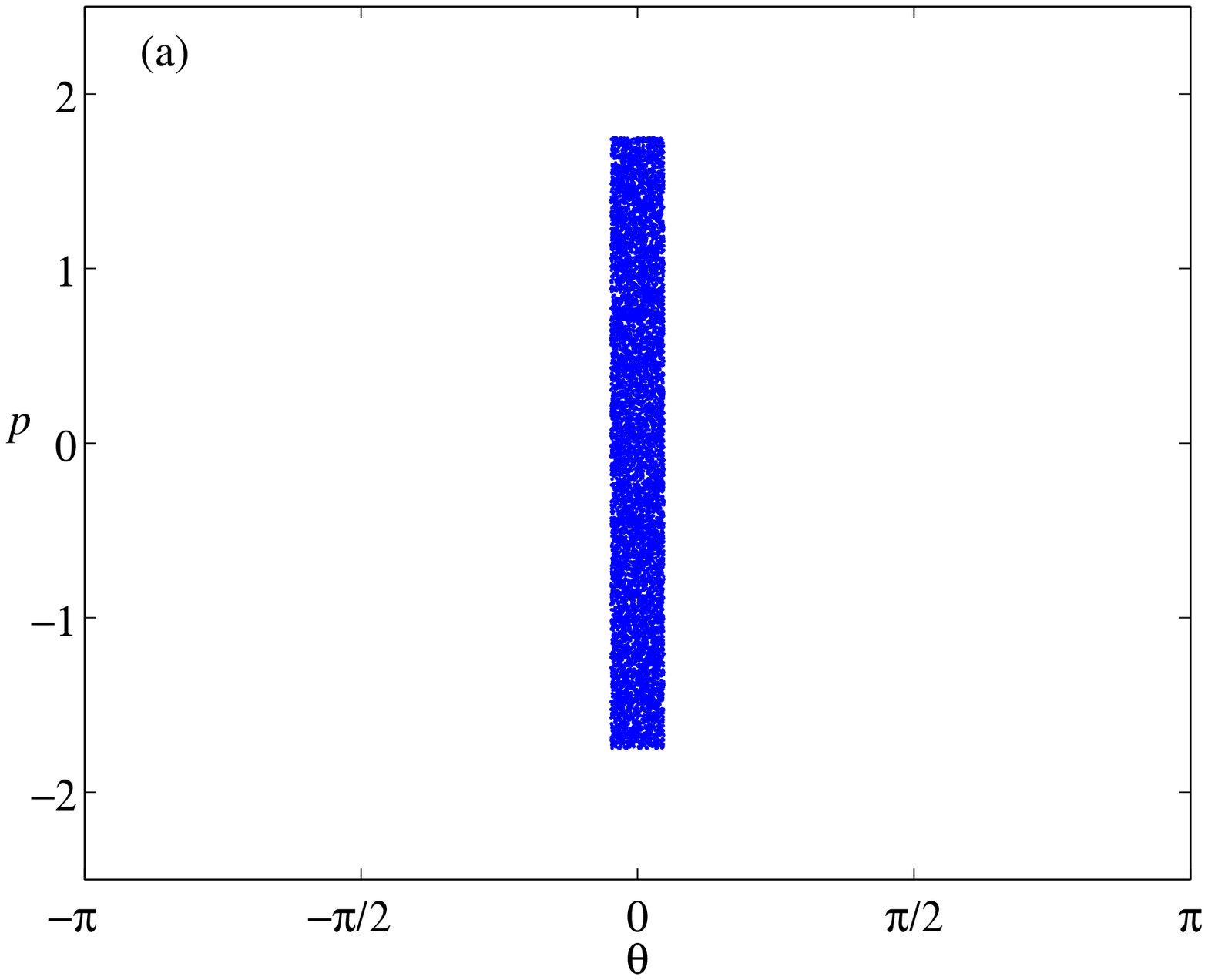}\hspace{1cm}
\includegraphics[scale=.3]{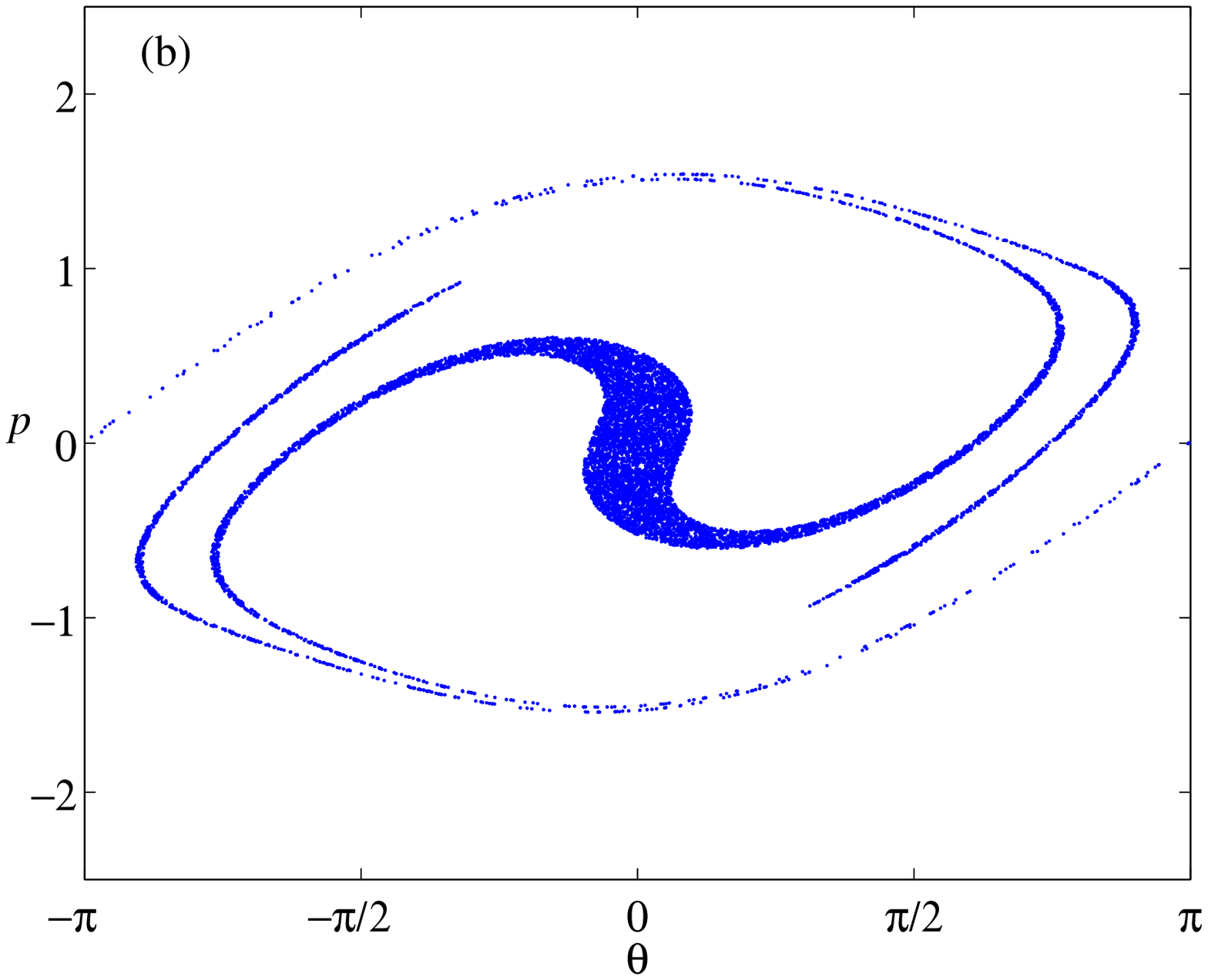}

\includegraphics[scale=.3]{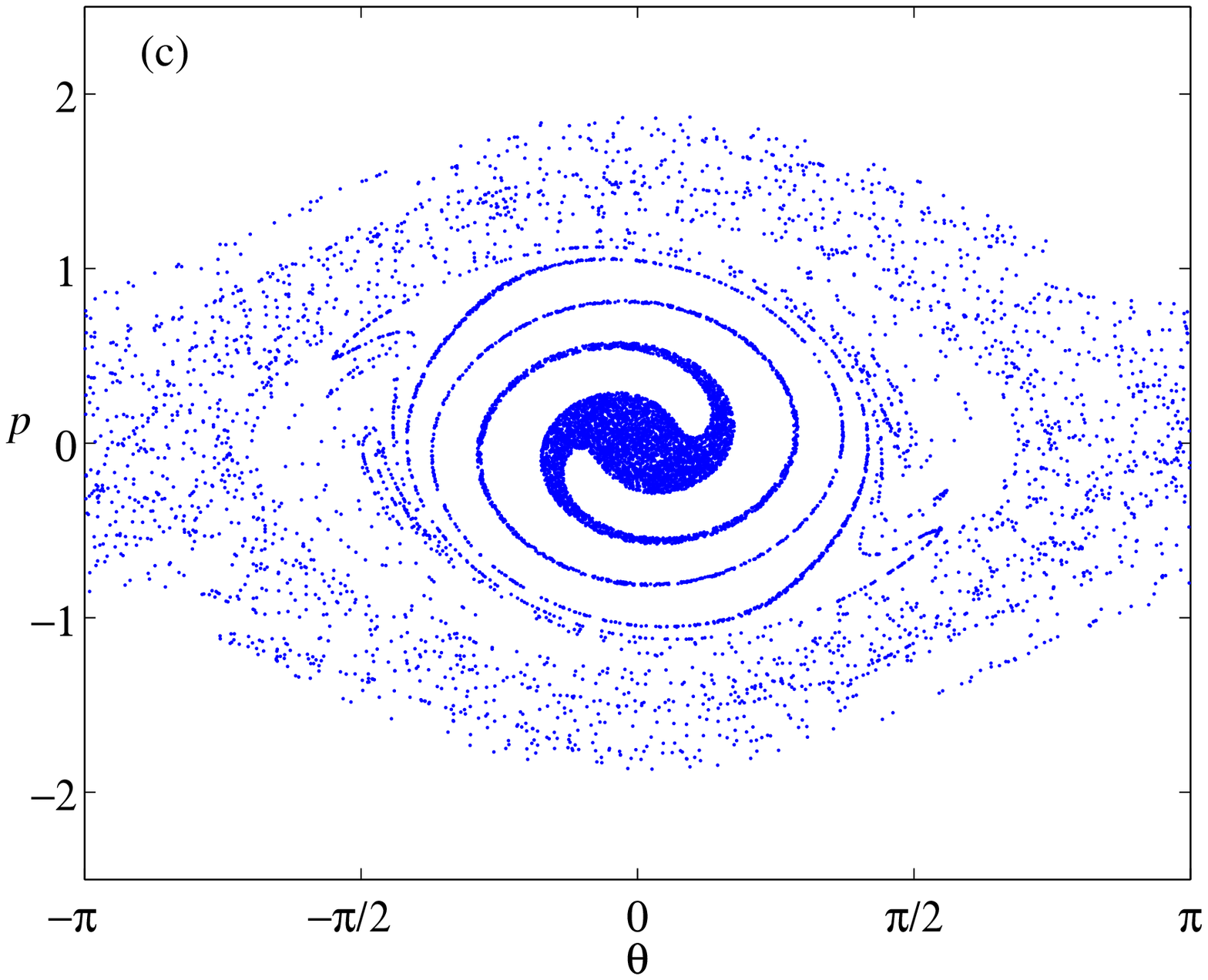}\hspace{1cm}
\includegraphics[scale=.3]{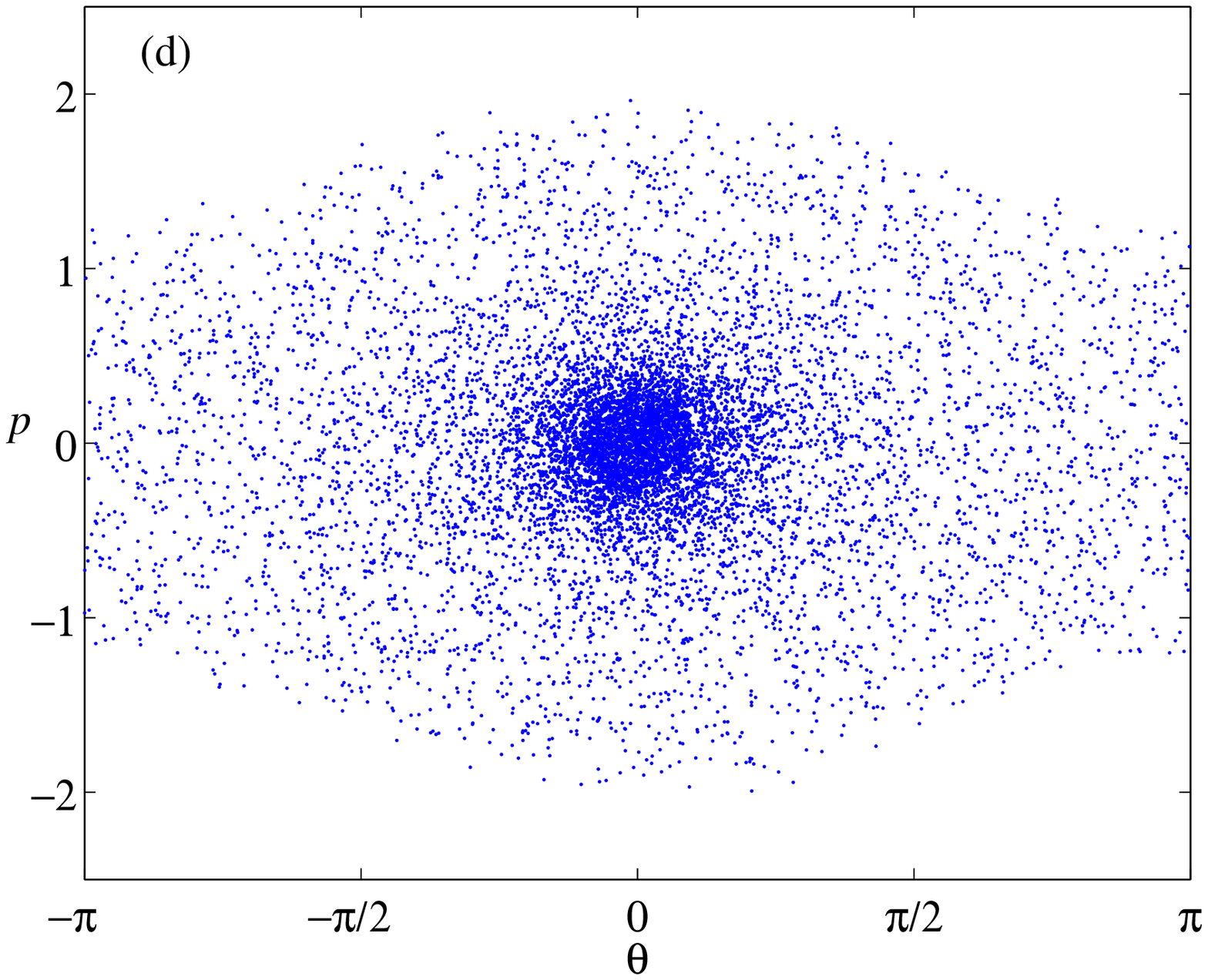}
\end{center}
\caption{Snapshots of the evolution of the $\mu$ phase space of the HMF model starting from a non-equilibrium magnetized initial condition of a waterbag form ($\theta_m=0.3$, $p_m=1.75$). The final QSS is magnetized: (a) $t=0.0$, (b) $t=100.0$, (c) $t=1000.0$, and (d) $t=10^5$.}
\label{fig7}
\end{figure}

\begin{figure}[!htb]
\begin{center}
\includegraphics[scale=.3]{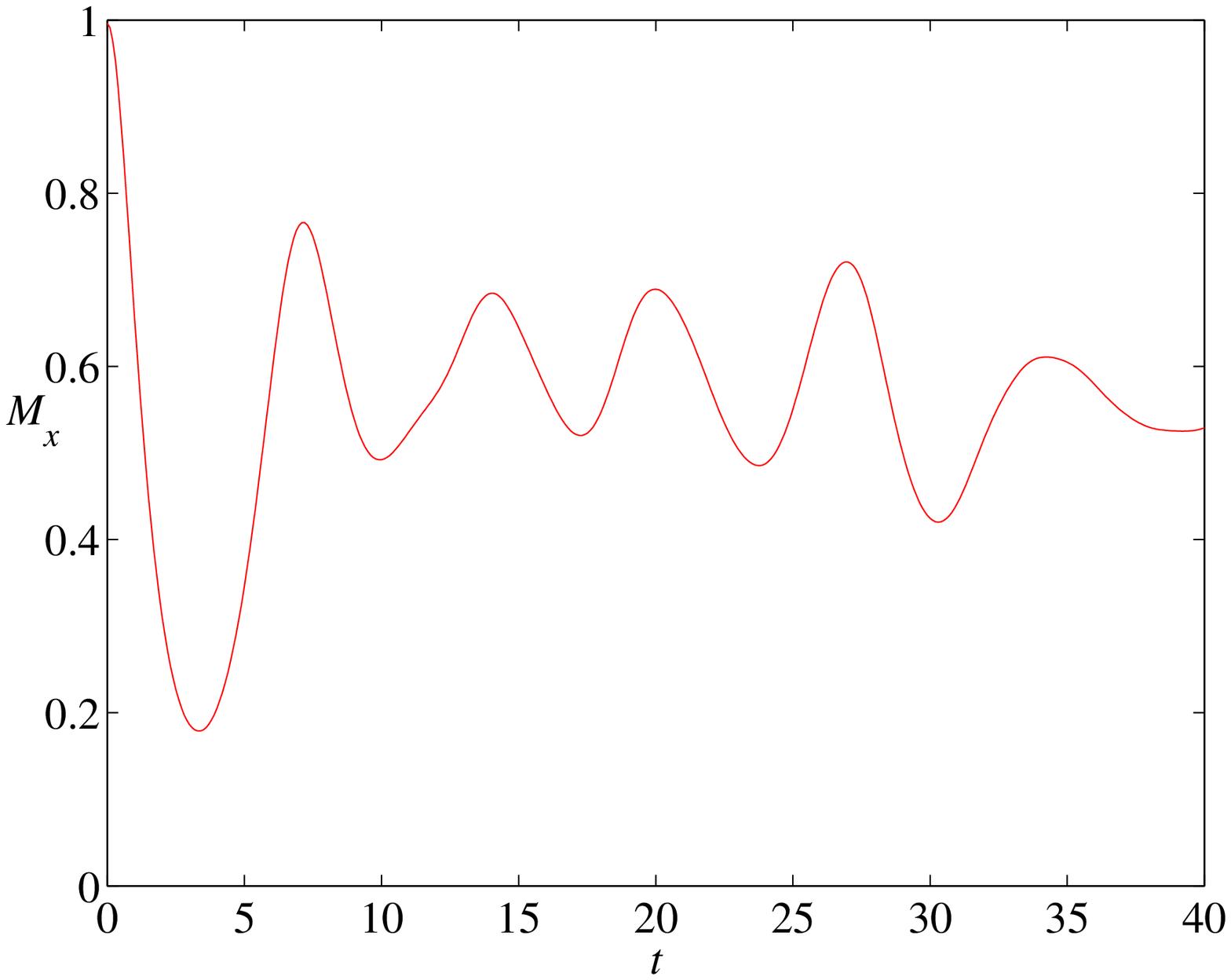}\hspace{1cm}
\end{center}
\caption{$M_x$ as a function of time for an initially magnetized state.  
The final QSS to which the system evolves is also magnetized.}
\label{fig10}
\end{figure}

\begin{figure}[!htb]
\begin{center}
\includegraphics[scale=.3]{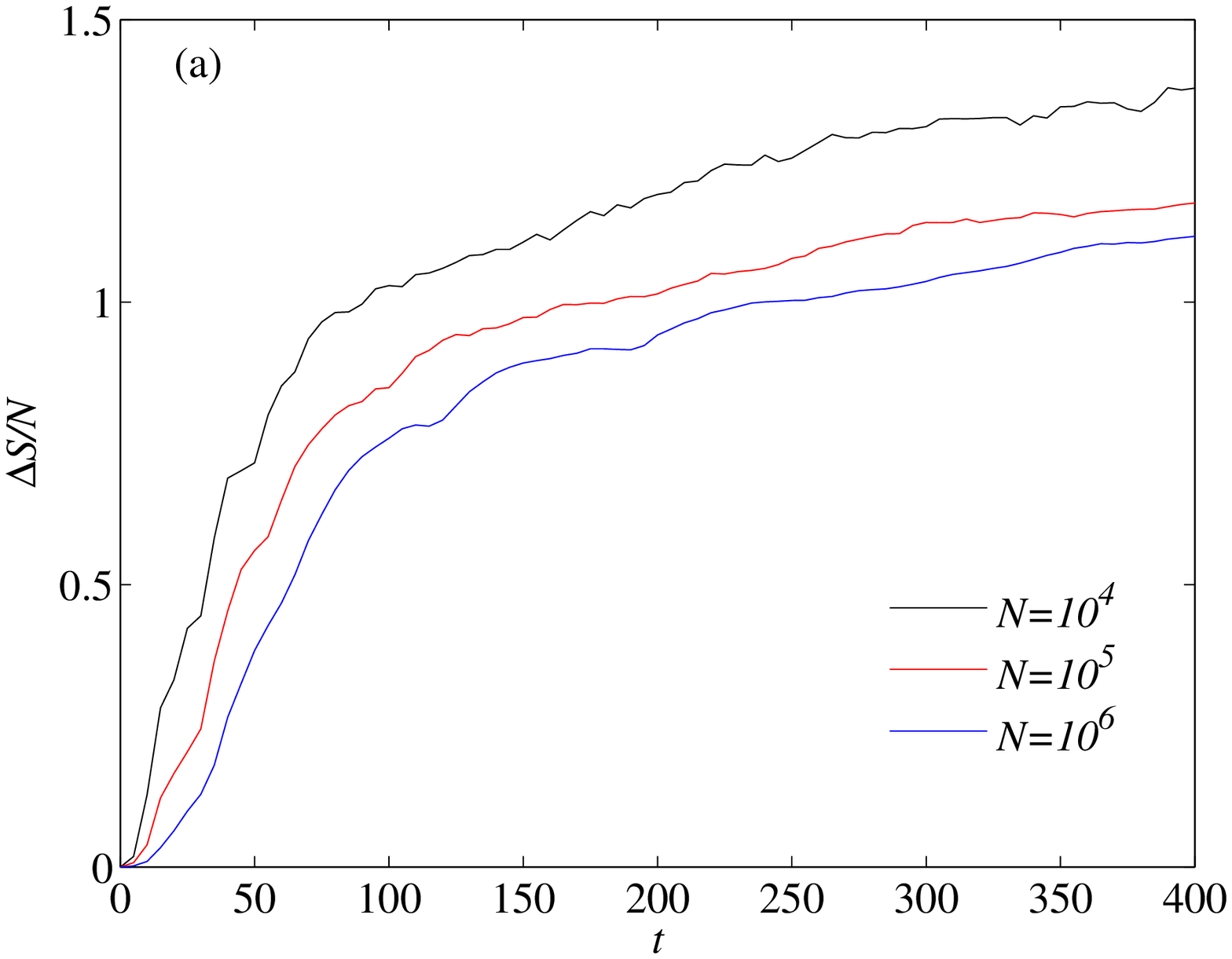}\hspace{1cm}
\includegraphics[scale=.3]{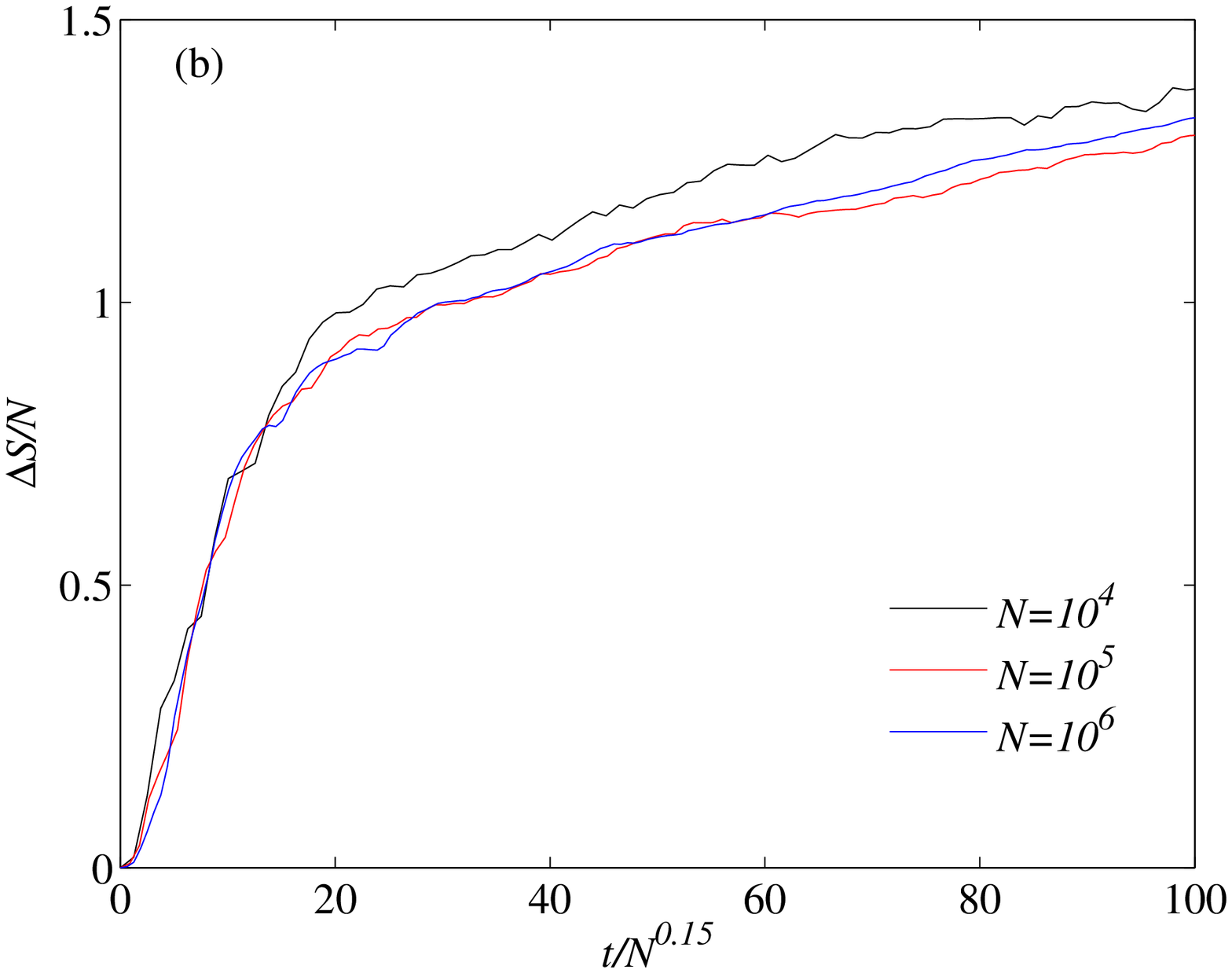}
\end{center}
\caption{Entropy production in the HMF model starting from a non-equilibrium magnetized initial condition of a waterbag form ($\theta_m=0.3$, $p_m=1.75$) which then relaxes to a magnetized QSS.  (a) dynamical time, (b) scaled time.}
\label{fig3}
\end{figure}

In our next example, see Fig. \ref{fig7}, we study an 
initially magnetized particle distribution which  relaxes
to a magnetized QSS, see Fig. \ref{fig10}.
Although the evolution of the phase space in this case is very different from the one that resulted
in an unmagnetized QSS,  the exponent $\alpha$ for the entropy production is the same: $\alpha=0.15$, see
Fig. \ref{fig3}.

\begin{figure}[!htb]
\begin{center}
\includegraphics[scale=.3]{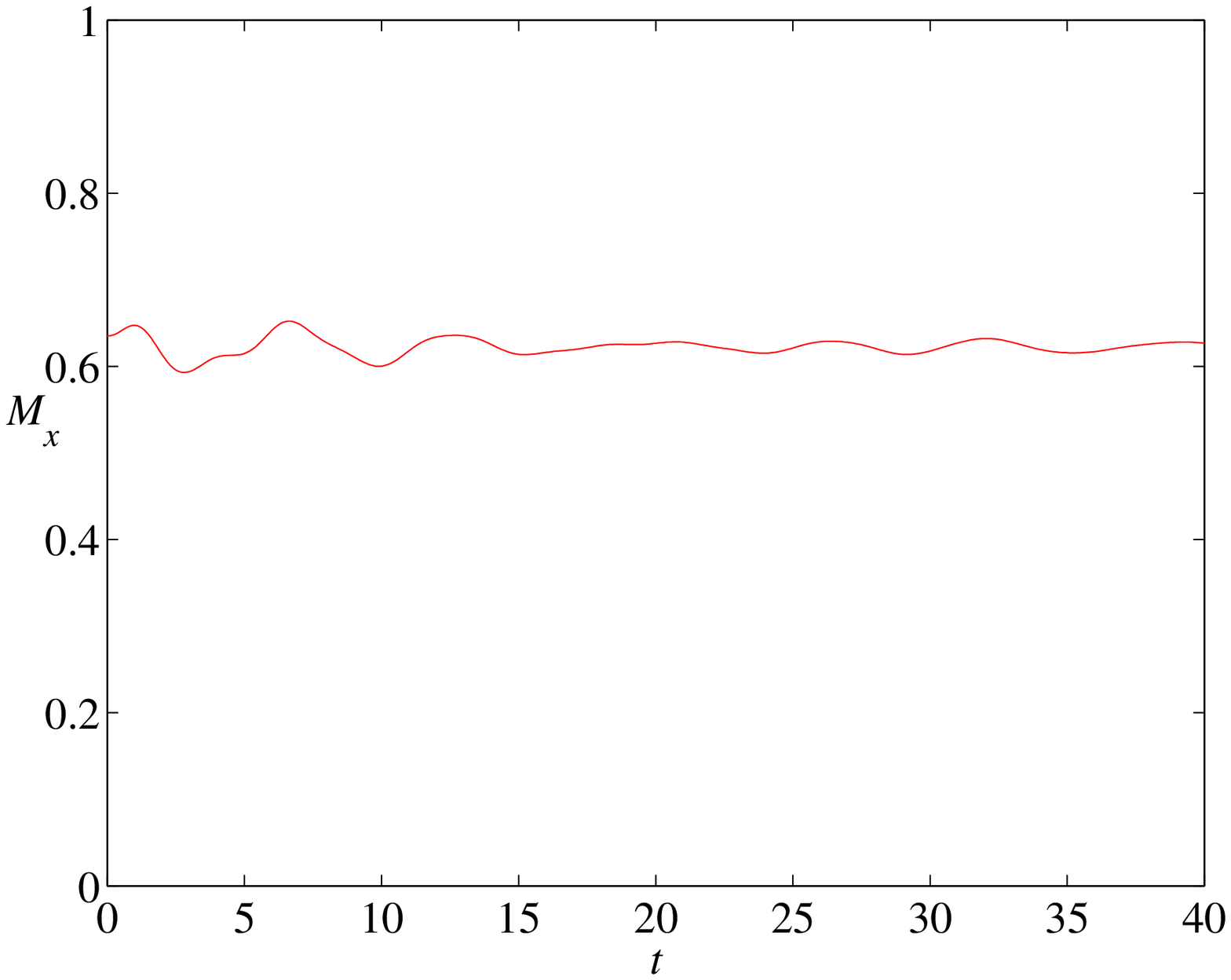}\hspace{1cm}
\end{center}
\caption{$M_x$ as a function of time for an initially magnetized state satisfying generalized virial condition (GVC). The oscillations of the magnetization are very small and $M_x$ in the final QSS is almost identical to the initial value.}
\label{fig11}
\end{figure}

\begin{figure}[!htb]
\begin{center}
\includegraphics[scale=.3]{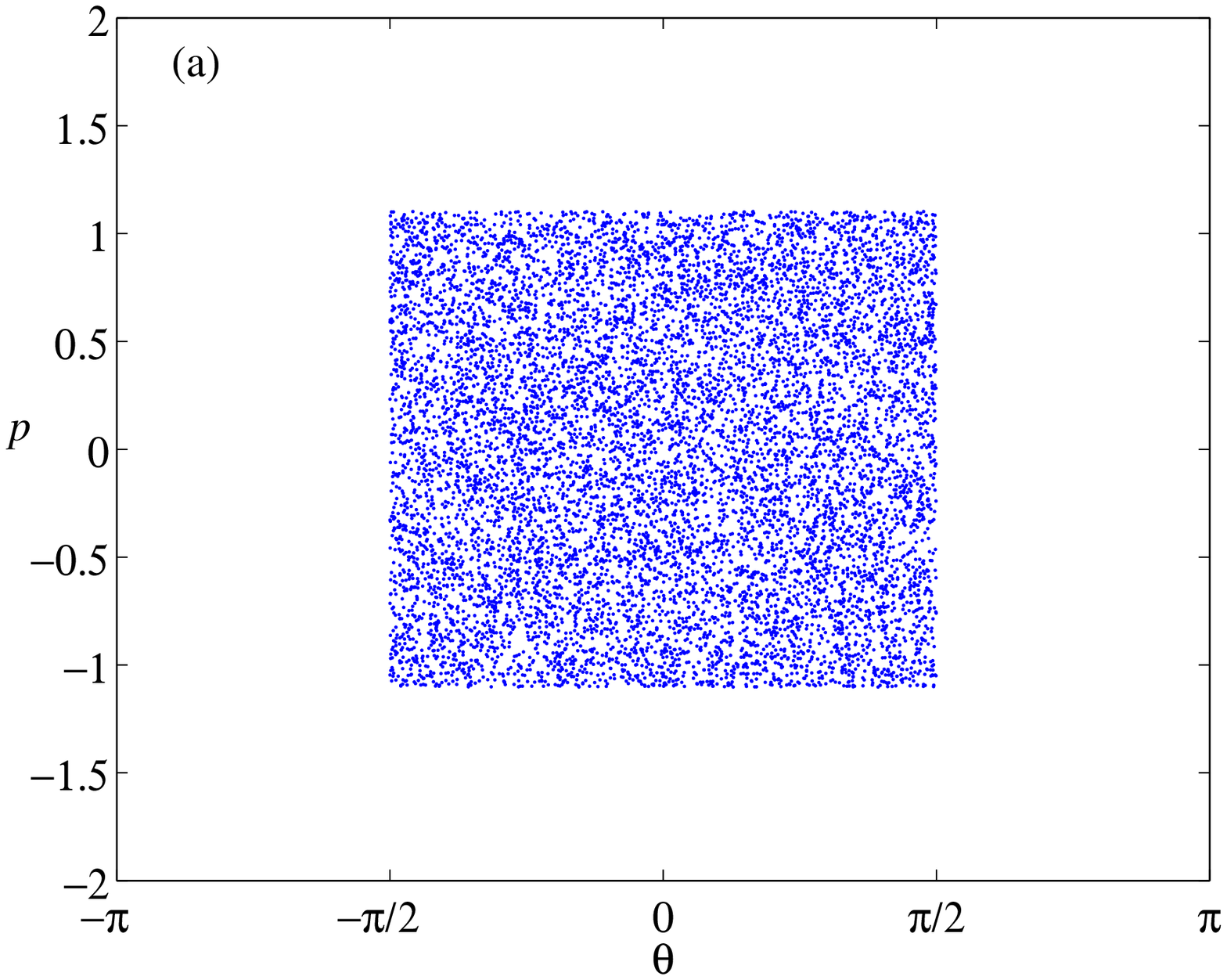}\hspace{1cm}
\includegraphics[scale=.3]{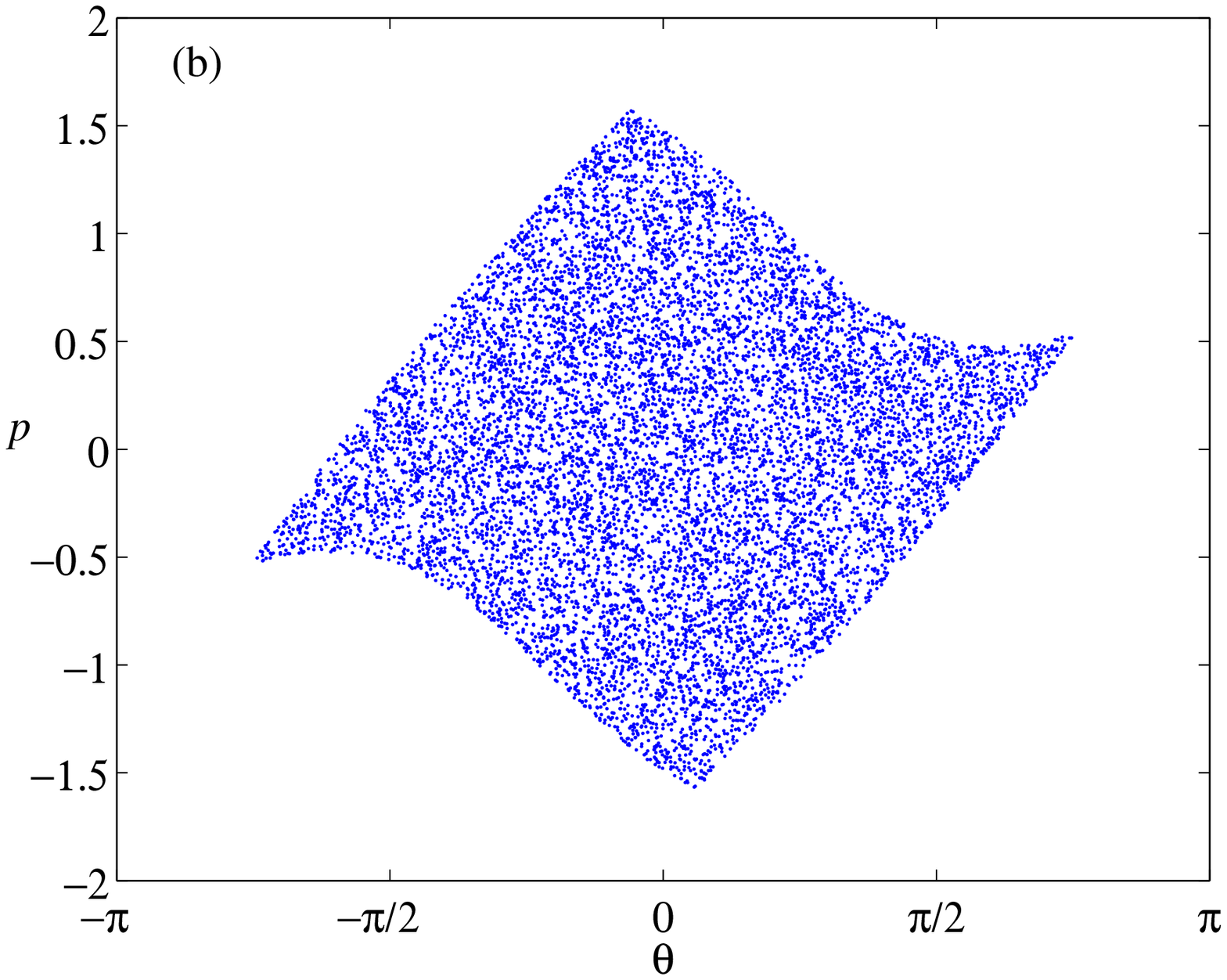}

\includegraphics[scale=.3]{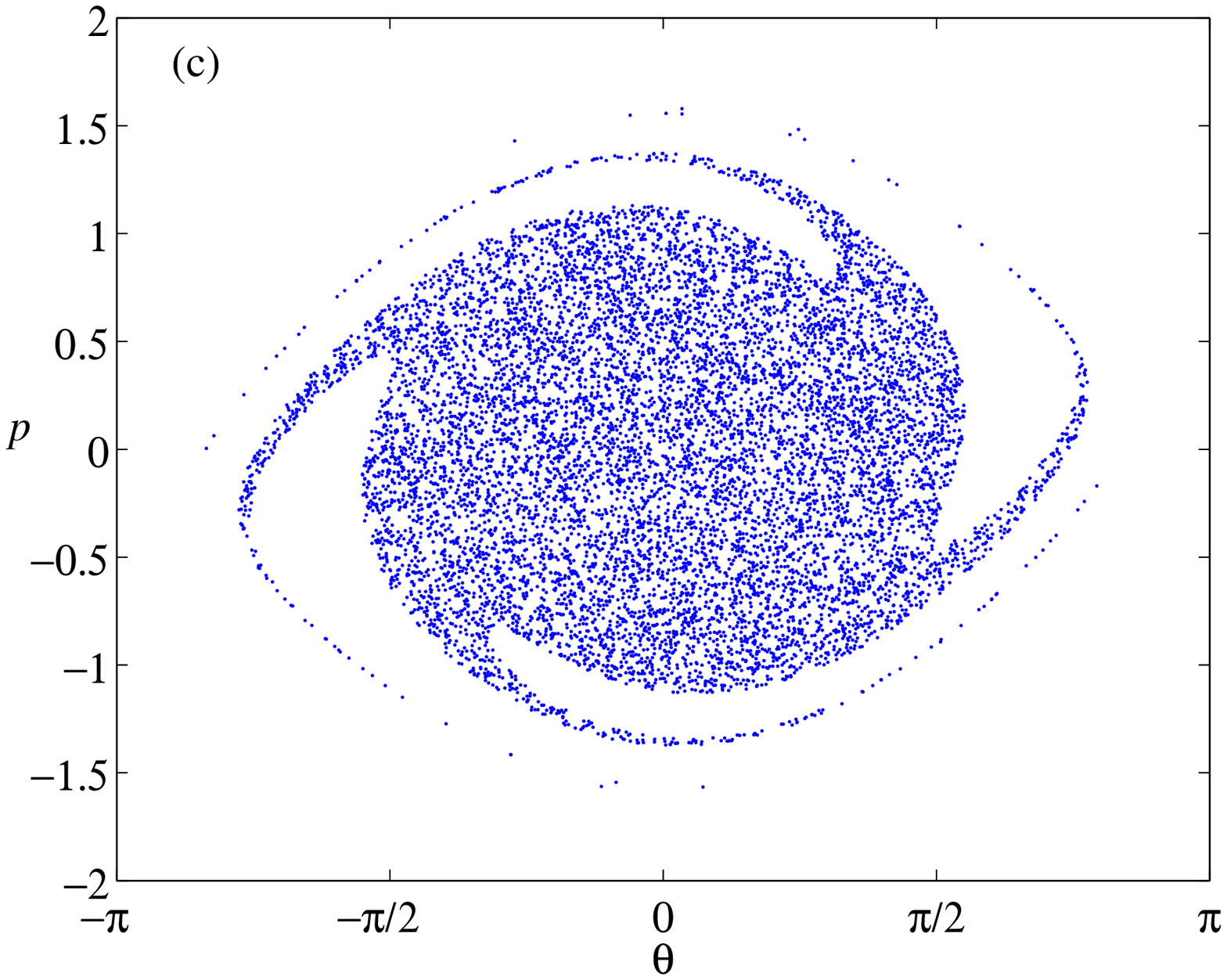}\hspace{1cm}
\includegraphics[scale=.3]{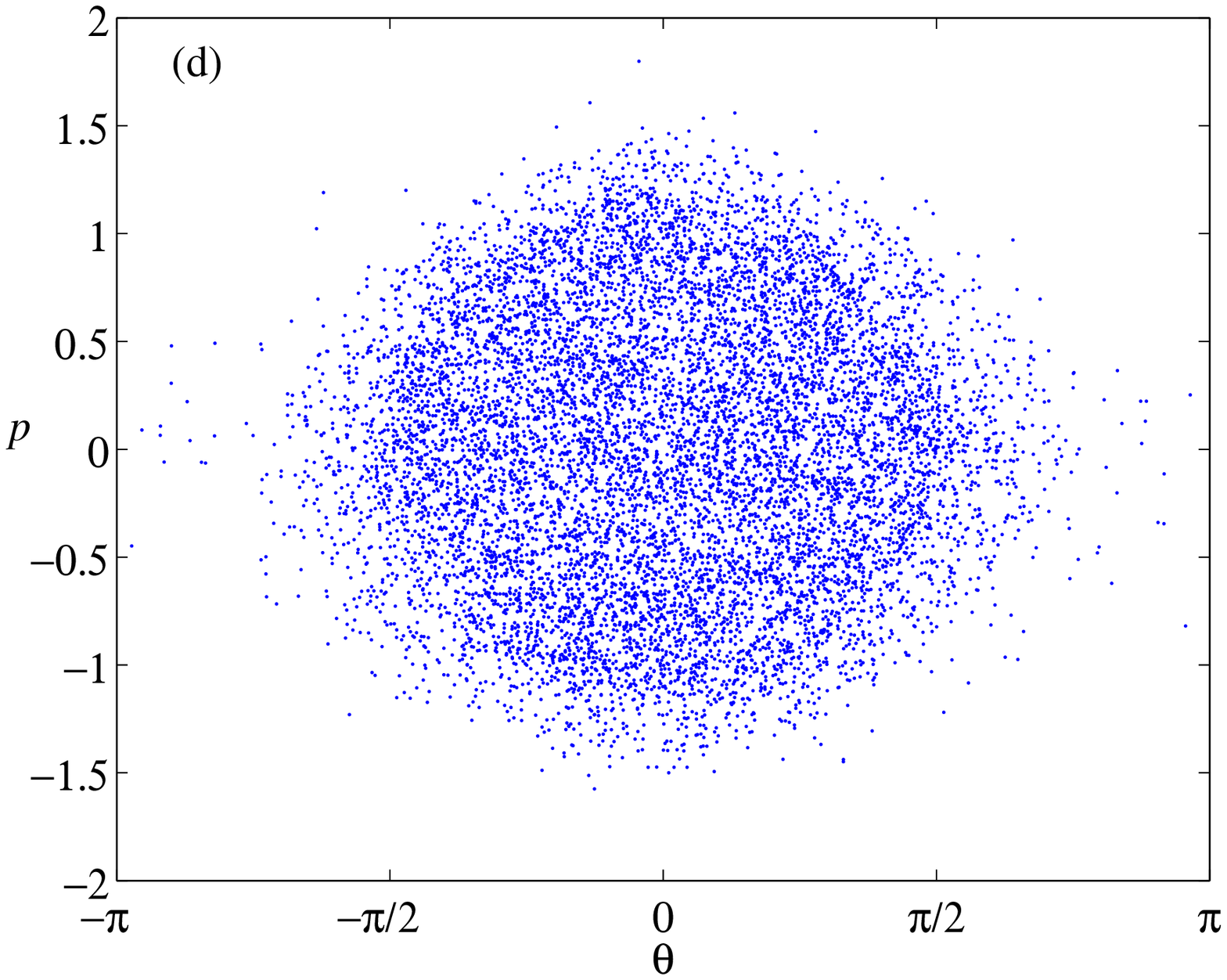}
\end{center}
\caption{Snapshots of the evolution of the $\mu$ phase space of the HMF model starting from an initial condition satisfying GVC ($\theta_m=\pi/2$, $p_m=1.10$). The final QSS is magnetized with $M_x$ almost identical to the initial value: (a) $t=0.0$, (b) $t=10.0$, (c) $t=100.0$, and (d) $t=10^4$.}
\label{fig8}
\end{figure}

\begin{figure}[!htb]
\begin{center}
\includegraphics[scale=.3]{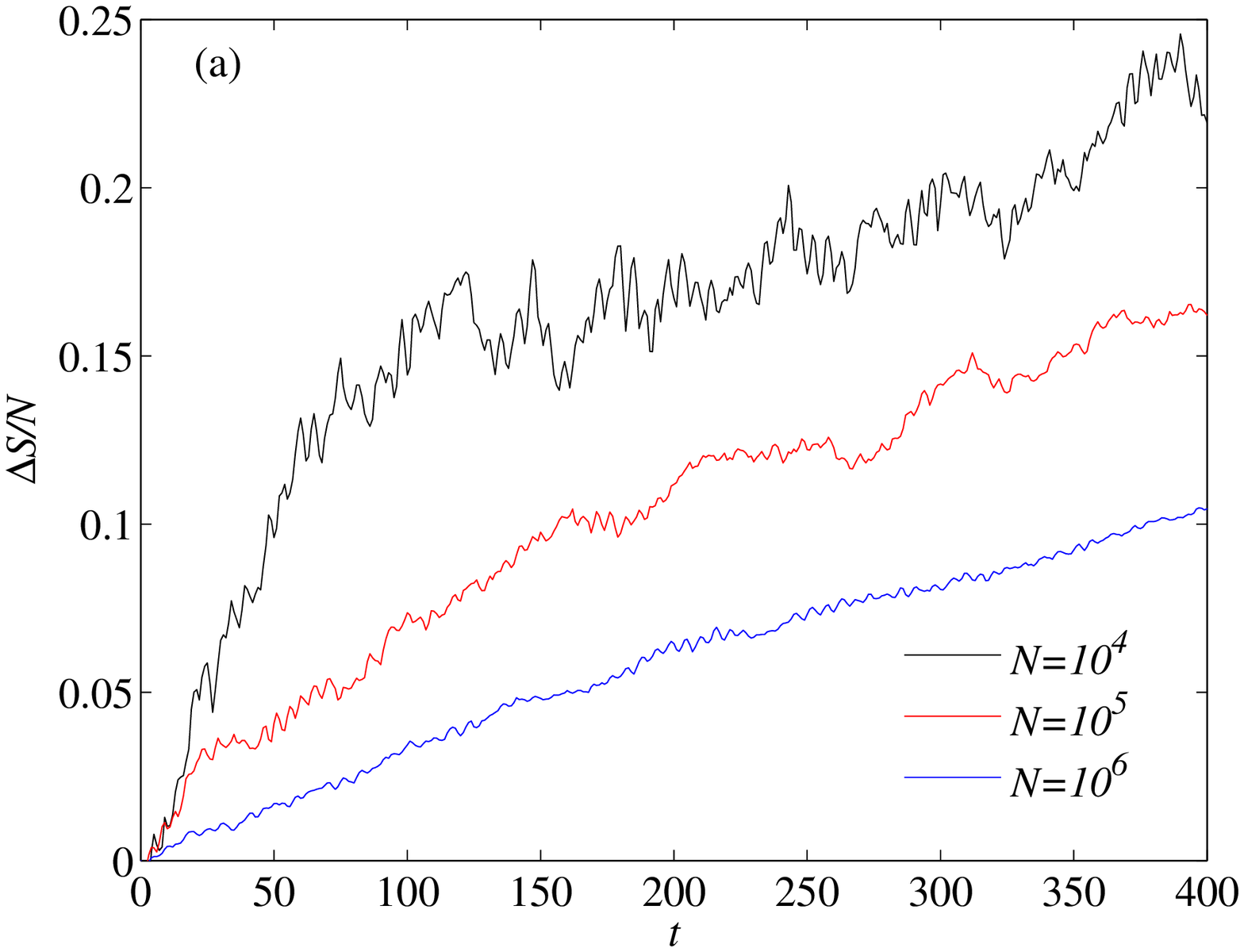}\hspace{1cm}
\includegraphics[scale=.3]{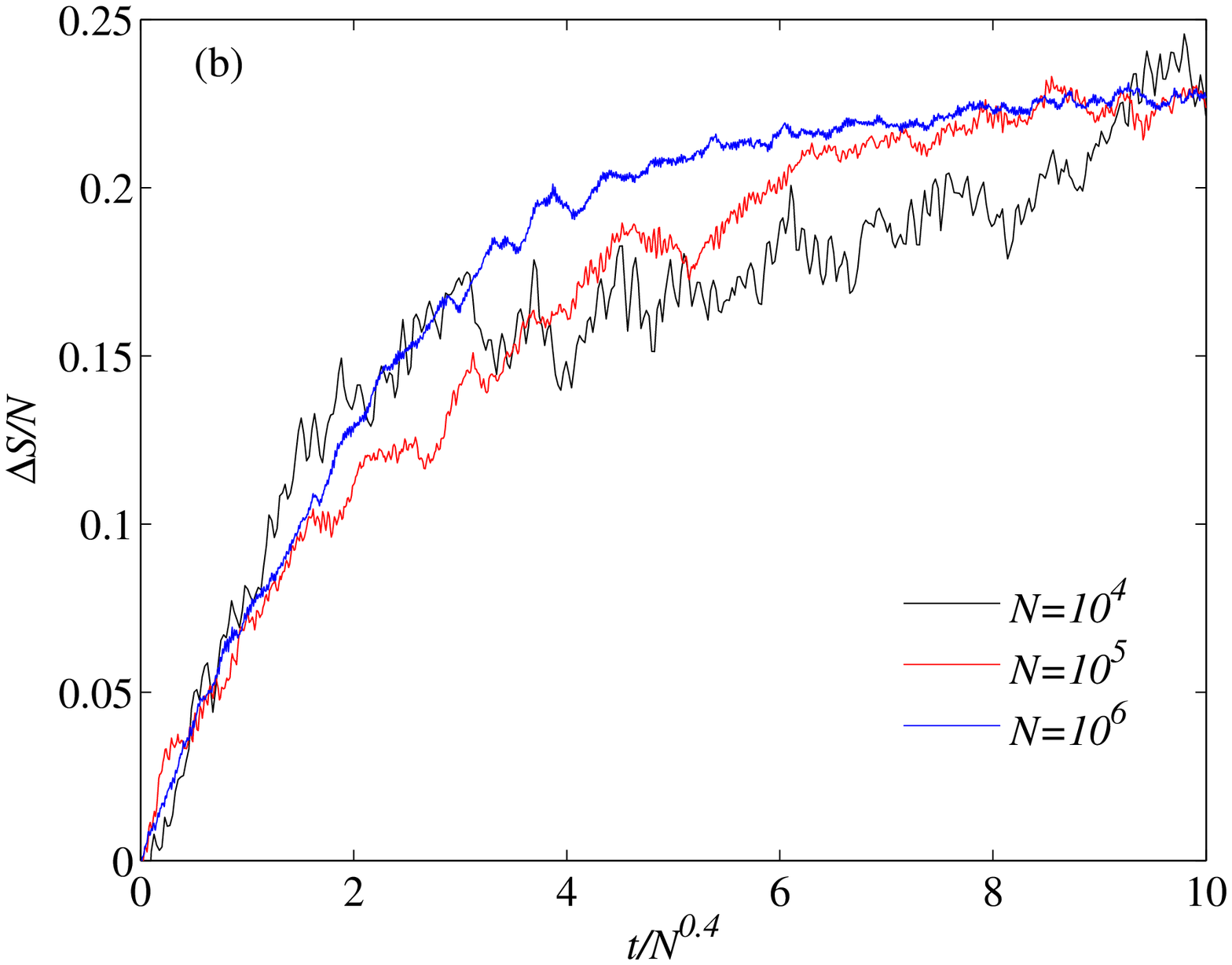}
\end{center}
\caption{Entropy production in the HMF model starting from an initial waterbag distribution satisfying the GVC ($\theta_m=\pi/2$, $p_m=1.10$): (a) dynamical time, (b) scaled time. }
\label{fig4}
\end{figure}

Our final example  is of an initial condition which is not ``too far" from
the final QSS. In this case, $p_m$ and $\theta_m$ of the initial waterbag distribution Eq. (\ref{wb}) are
adjusted so that the generalized virial condition (GVC) is satisfied~\cite{BeTe12}.  Under this condition the parametric resonances
are suppressed and the dynamical evolution towards the final QSS is adiabatic~\cite{RiBe14}.  The magnetization
shows only very small oscillations about the initial value and the final magnetization of the QSS is almost identical to that of the initial distribution, see Fig. \ref{fig11}. 
The snapshots of the evolution of $\mu$ phase space are shown in Fig. \ref{fig8}.
The entropy production is shown in Fig. \ref{fig4}.  Once again the data can be collapsed by rescaling 
time with $N^\alpha$, however, unlike for non-GVC initial conditions we now find $\alpha=0.4$, showing that
the mixing of the phase space is significantly slower.

\section{Conclusions}

We have explored entropy production in systems with long range interactions.  Differently from systems in which particles interact through short range potentials, entropy production for long range interacting systems can be calculated explicitly.  This is possible because the probability distribution function for long range systems in the thermodynamic limit reduces to the product of one particle distribution functions, and the full $2dN$ dimensional $\Gamma$ phase space collapses to $2d$ dimensional $\mu$ phase space.  Under these conditions the coarse grained Gibbs entropy can be obtained explicitly using non-parametric entropy estimator.  We discover that the entropy production takes place on a time scale which grows with the number of particles as $N^\alpha$. 
In the case of the HMF model we find that if the particle dynamics has parametric resonances, the exponent is $\alpha \approx 0.15$. On the other hand if dynamics is adiabatic  -- which is the case for the initial particle distributions that satisfy GVC -- then $\alpha=0.4$, which is close to the exponent found for non-interacting particles.  It will be interesting to explore how universal are these exponents by studying other long range systems, such as magnetically confined plasmas~\cite{LePa08} or self-gravitating clusters~\cite{YaMi03,JoWo11,TaLe10,SiRo16}.  The fact that the paramagnetic resonances and chaotic dynamics diminish significantly the entropy production time suggests that for short range interacting systems, for which dynamics is highly non-linear and chaotic, the 
exponent $\alpha \rightarrow 0$, and the entropy production will take place on a microscopic time scale even in the thermodynamic limit.

This work was partially supported by CNPq, Brazil, and by the US-AFOSR under Grant No. FA9550-16-1-0280. The authors are grateful to Leandro Beraldo e Silva for 
interesting conversations.

\end{document}